\documentclass[12pt]{article}
\usepackage{amsmath,amsfonts,amsthm,bm} 
\usepackage{graphicx,psfrag,epsf}
\usepackage{enumerate}
\usepackage{natbib}

\usepackage{url}
\usepackage{chngcntr}
\usepackage{dsfont }
\usepackage[dvipsnames]{xcolor}
\usepackage[utf8]{inputenc}
\usepackage{relsize}
\usepackage{amssymb }
\usepackage{enumitem}
\usepackage[english]{babel}	
\usepackage{upgreek }
\usepackage{multirow}
\usepackage[toc,page]{appendix}
\usepackage{boldline}
\usepackage{caption, subcaption}

\newcommand{\mf}{\mathbf}
\newcommand{\bs}{\boldsymbol}
\newcommand{\ms}{\mathds}

\newcommand{\argmin}{\arg\!\min}
\newcommand{\argmax}{\arg\!\max}

\newcommand{\ve}[1]{\mbox{\boldmath ${#1}$}}

\usepackage{authblk}

  \title{Simultaneous Registration and Clustering for Multi-dimensional Functional Data}
  \author[1]{Pengcheng Zeng}
  \author[1]{Jian Qing Shi \thanks{Corresponding author, email: j.q.shi@ncl.ac.uk}}
  \author[2]{Won-Seok Kim}
    
\affil[1]{School of Mathematics, Statistics and Physics, Newcastle University, UK}
\affil[2]{Department of Rehabilitation Medicine, Seoul National University, Korea}
\date{\today}

\begin{document}
	
\maketitle  	
	\bigskip
	\begin{abstract}
		The clustering for functional data with misaligned problems has drawn much attention in the last decade. Most methods do the clustering after those functional data being registered and there has been little research using both functional and scalar variables. In this paper, we propose a simultaneous registration and clustering (SRC) model via  two-level models,   allowing the use of both types of variables and also allowing simultaneous registration and clustering.  For the data collected from subjects in different unknown groups, a Gaussian process functional regression model with time warping is used as the first level model;  an allocation model depending on scalar variables is used as the second level model providing further information over the groups. The former carries out  registration and modeling for the multi-dimensional functional data (2D or 3D curves) at the same time. This methodology is implemented using an EM algorithm, and  is examined on both simulated data and real data.
		
	\end{abstract}
	
	\noindent%
	
	\section{Introduction}
	\label{sec:intro}
	Functional data analysis (FDA) has many applications in almost every branch of science, such as engineering, medicine, geology and biology. Basically, it aims at dealing with the analysis of data in the form of images, curves and shapes. But it always comes along with some challenges, like warping time (or misaligned), observation noise and infinite-dimensionality of function spaces. Among those challenges, data registration plays an important role in terms of preprocessing  \citep{Ram05}. In curves, the lateral displacement termed as phase variation, as opposed to amplitude variation in curve height, has drawn much attention.  It is necessary to remove the phase variation from the amplitude in a desirable fashion, since it always increase data variance, distorts principal components and make the underlying data structures unclear \citep{Mar15}. Time warping function is then introduced for this purpose. If we denote the system time or internal time scale as $t$, the underlying time process shared by all the observations, the functional relationship $g^{-1}(t)$, called time-warping function,  represent the clock time or individual-specific time scale, varying on from another. The choice of warping function is dependent on the particular application context and the methods of estimating $g^{-1}(t)$ differ. Generally, among those strategies, are seeking various metric for $g^{-1}(t)$, such as $L^{2}$ distance \citep{Ger04, Tan08}, similarity index \citep{San09} and Fisher-Rao metric \citep{Sri11b}, as well as modeling $g^{-1}(t)$ directly or indirectly and estimating it through MLE \citep{Lars:2016} or Bayesian inference \citep{Che16}. Most of the existing methods target the registration of one dimensional functional data and only a few of them, e.g. \cite{San09}, \cite{Sri11a} and \cite{Che16}, can be applied to multi-dimensional case.
	
	On the other hand, the clustering for functional data  has  been receiving much attention. The aim is to group a set of data such that data within groups (clusters) are more similar than across groups with respect to a metric. It is often used as a preliminary step for data exploration by identifying particular patterns to provide the user with convenient interpretation.  It is generally a difficult task due to the lack of distances or estimation from noise data and a definition for the probability of  a functional variable. Loads of different approaches have been proposed. For instance, the nonparametric methods defining specific distance or dissimilarities,  like hierarchical clustering \citep{Hart78, Fer06} and  k-means algorithm \citep{Hart78, Iev13, Tar03, Tok07}, and the distribution-based clustering, such as modeling principal components \citep{Del10, Bou11} or basis expansion coefficients \citep{Jam03, Sam11}) with mixture Gaussian distributions, or curves themselves by mixture Gaussian process \citep{Shi:05, Shi:08}.
	
	\begin{figure}	
		\centering
		\begin{subfigure}[t]{0.39\textwidth}
			\centering
			\includegraphics[width= \textwidth]{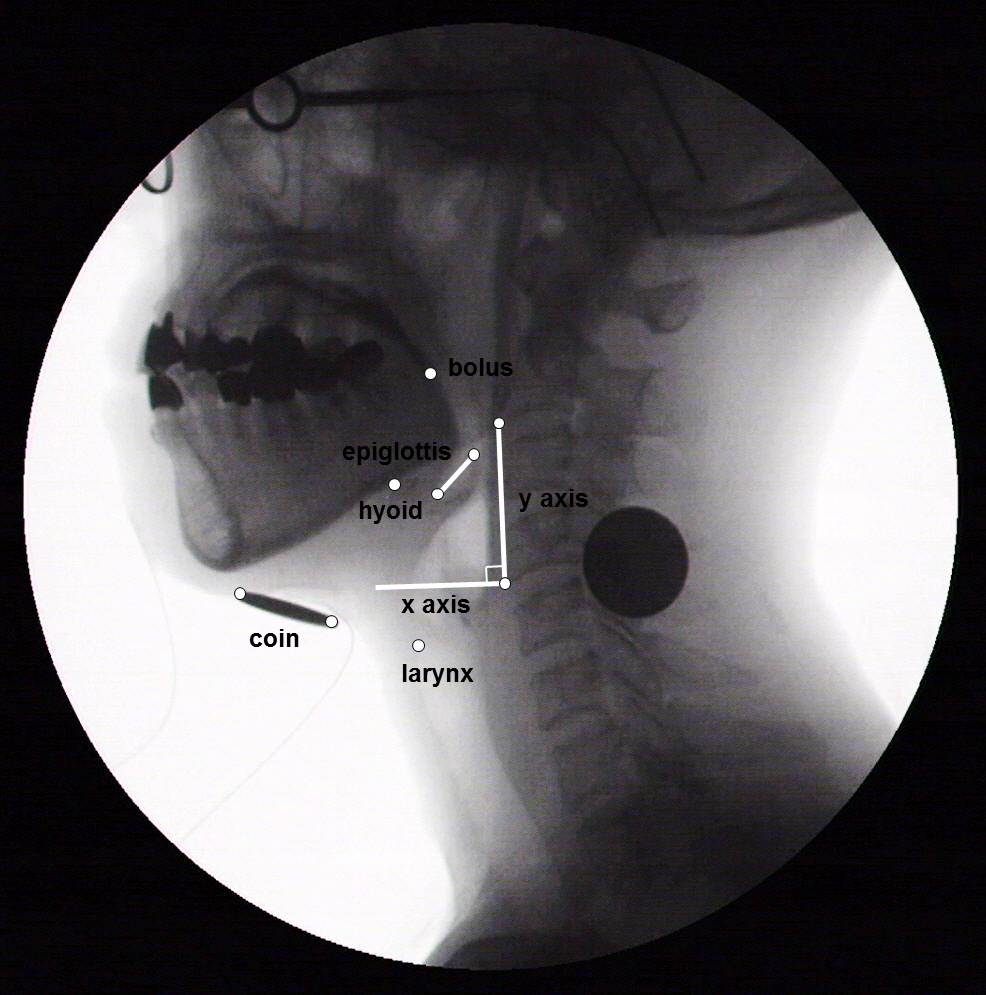}
			\caption{}		
		\end{subfigure}
		\quad
		\begin{subfigure}[t]{0.46\textwidth}
			\centering
			\includegraphics[width= \textwidth]{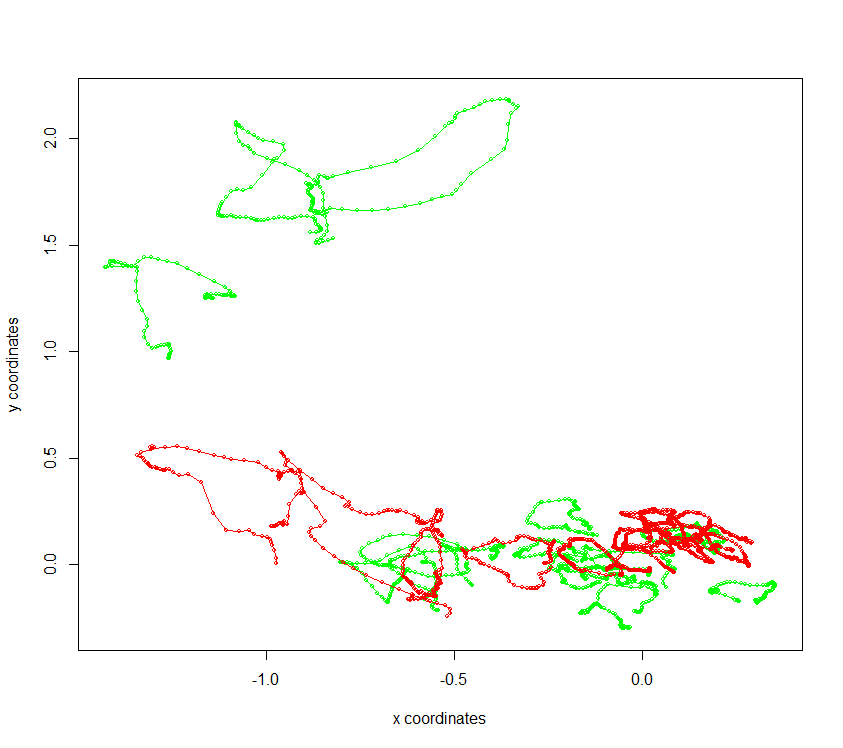}
			\caption{}		
		\end{subfigure}
		\caption{The motion data of  hyoid bone. ($\mf{a}$) One X-ray image showing the location of hyoid bone which will move forward and backward to form one 2D curve during swallowing, as shown in ($\mf{b}$). ($\mf{b}$) 30 trajectories of hyoid bone motion from 15 normal people (curves in green) and 15 patients after stoke (curves in red).} \label{fig:realdata}
	\end{figure}
	
	In our study of the motion analysis of hyoid bone observed from X-ray video clips, the trajectories can be thought of as 2D functional data (see Figure \ref{fig:realdata}, the background of the data will be discussed in Section \ref{sec:num_ana}). There exists obvious misaligned problems for those curves in both vertical  and horizontal variation.  Usually, curve alignment is performed as a preprocessing technique and the clustering is conducted afterwards. This way is not efficient, since a subject belonging to which cluster is closely related to how it unfolds its progression pace. Another challenging problem for this study is that the heterogeneity of regression relationships among different groups. It consists in both the potential time warping for curves corresponding to the subjects and the subjects' scalar variables such as initial level of disease, gender, age and the characteristics of those trajectories themselves, like motion time, average speed and range of motion. Therefore, simultaneous curve registration and clustering by considering all those factors  seems to be a better way for modeling the functional data.
	
	There are some research work on handling the similar problems. For instance, \cite{Liu:09} developed a framework that allows for simultaneously aligning and clustering k-centers functional data. But their model did not use any subject specific information (scalar variables)  and assumed the heterogeneity among groups just depends on the curves themselves; similar idea is also used in  $k$-means alignment for curve clustering by \cite{San10}. On the other hand, \cite{Shi:08} proposed a hierarchical mixture of Gaussian process (GP) functional regression models with an allocation model to do curve prediction and clustering. They used the functional covariates to reconstruct the response curve and the personal scalar variables, such as height and gender, to deal with heterogeneity of the regression relationships among different groups. However, their method did not consider the misaligned problem. In addition, most related models are limited to one dimensional curve.
	
	The purpose of our study is to address the above problems by constructing one hierarchical mixture of models for the sake of simultaneous curve registration and clustering. In the first level,  we assume a simultaneous registration and  functional regression model. And an allocation model with scalar variables is introduced as the second level model. Non-linear functional random effects are modeled by a GP in the first level model to address the heterogeneity among subjects.  This also allows to estimate  subject-specific warping function. The main contributions of this paper are: (i) simultaneously carrying out registration and modeling for multi-dimensional functional data (2D or 3D curves) allowing variation among subjects, and (ii) the use of both functional and scalar covariates while conducting  clustering.
	
	The paper is organized as follows. Section \ref{subsec:mode} defines simultaneous registration and clustering (SRC) models via two-level models. We discuss the  estimation and the details of implementation  in Section \ref{subsec:esti} and Section \ref{subsec:impl} respectively.  The problem of model selection and the  related methods are discussed in Section \ref{subsec:mod_sel}. Section \ref{sec:num_ana} presents a number of examples with simulated data and real data. A short summary and discussion are given in Section \ref{sec:dis}.
	
	\section{The simultaneous registration and clustering method}
	\label{sec:meth}
	Suppose there are $N$ subjects coming from $K$ different groups, $\ve{x}_{1}, \ve{x}_{2},\dots, \ve{x}_{N}$  being the observations of 2D continuous curves, where $\ve{x}_{i} = (\ve{x}_{1i}(t), \ve{x}_{2i}(t))^{\intercal}$, $\ve{x}_{1i}(t)$ and $\ve{x}_{2i}(t)$ are the corresponding $x$-coordinates and $y$-coordinates of $\ve{x}_{i}$. Let $\ve{v}_{1},\dots, \ve{v}_{N}$ be the observed scalar variables. 
	Suppose there are $n_{i}$ time points on which the $i$-th curve is measured. The data set is
	\[
	D = \{(\ve{x}_{i}(t_{ij}), \ve{v}_{i}); i = 1,\dots, N; j = 1,\dots, n_{i}\}.
	\]
	We introduce a latent indicator variable $\ve{z}_{i} = (z_{i1}, \dots, z_{iK})^{\intercal}$  for the $i$-th subject where $z_{ik}$ takes value $1$ if they are in the $k$-th group and $0$ otherwise. 
	\subsection{The model}
	\label{subsec:mode}
	
	In our study of 2D curves,  we will use the preprocessing procedure Generalized Procrustes Analysis (GPA) \citep{Gower:1975} to address part of registration problems  in advance except warping.  Conventionally,  most methods tried to complete all the registration problems including warping before clustering.  This is not the best way since different warping functions may need to be used in different clusters, yet we have no such information before clustering, and heterogeneity among different subjects should also be considered.
	Thus, a hierarchical structure defined by two levels of models is proposed. 
	
	We start with the first level model for the continuous curve as follows
	\begin{equation}\label{eq:mixt}
	x_{ai}(t)|_{z_{ki}=1} = (\tau_{ak}\circ g_{ki})(t) + r_{aki}(t) + \epsilon_{ai}(t), \hspace{5mm} i = 1,\dots,N,
	\end{equation}
	where $a = 1$ or 2  represents $x$- or $y$-coordinates of $\ve{x}_{i}(t)$. The item $(\tau_{ak}\circ g_{ki})$ denotes functional composition: $(\tau \circ g)(t) = \tau(g(t))$, where $g_{ki}(t)$ is the inverse of a warping function. $\tau_{ak}(\cdot)$ is a fixed but unknown nonlinear mean curve, which can be approximated by a set of basis functions, the details will be given in the next subsection. The variation among different subjects is modeled by a non-linear functional random-effects, $r_{aki}(t)$, by a Gaussian process with zero-mean and a parametric covariance function $S$ \citep{Shi:12}. The error item $\epsilon_{ai}(t)$ is assumed to be Gaussian white noise with variance $\sigma^{2}$. 
	
	Following the previous discussion, we need to use different warping function in different cluster, and we also need to consider the variation among different subjects, and thus, we allow warping function depending on $k$ and $i$. Further, we assume 
	\[
	g_{ki}(t) = t + w_{k}(t) + w_{ki}(t),
	\]
	where $w_{k}(t)$  is the fixed part and $w_{ki}(t)$ is the random part in terms of different subjects. Instead of making assumption for curves $w_{k}(t)$ and $w_{ki}(t)$, we first discretize them by a set of fixed parameters, for example, by $\ve{w}_{k} = (w_{k}(t_{1}), \dots, w_{k}(t_{n_{v}}))$  and $\ve{w}_{ki} = (w_{ki}(t_{1}), \dots, w_{ki}(t_{n_{w}}))$ respectively. We then model $\ve{w}_{ki}$ by a Gaussian distribution with zero mean and a parametric covariance function $\ve H$. We can also define the warping function as simple as a horizontal shift, i.e. $g_{ki}(t) = t + b_{ki}$ or a linear stretch of the curves, i.e. $g_{ki}(t) = (1 + b_{ki})t + c_{ki}$, where $b_{ki}$ and $c_{ki}$ are both one dimensional unknown parameter. Those linear warping functions have been examined by others \citep{Liu:09, San10}.
	
	We define a logistic allocation model in the second level model for the latent indicator variable in the form
	\begin{equation}\label{eq:allo}
	p(z_{ki} = 1) = \pi_{ki} = \frac{\text{exp}\{\ve{v}^{\intercal}_{i}\ve{\beta}_{k}\}}{1 + \sum_{j =1}^{K-1}\text{exp}\{\ve{v}^{\intercal}_{i}\ve{\beta}_{j}\}}, \hspace{5mm} i = 1, \dots, N; \hspace{5mm} k = 1,\dots, K-1,
	\end{equation}
	with $p(z_{Ki} = 1) = \pi_{Ki} = 1 - \sum_{l=1}^{K-1}\pi_{li}$, where $\{\bs{\beta}_{k}, k = 1, \dots, K-1\}$ are unknown parameters to be estimated. We can also replace model (\ref{eq:allo}) by other models, e.g. Potts model \citep{Gre:00}. The information of scalar variables is integrated with functional variables via the two-level models (\ref{eq:mixt}) and (\ref{eq:allo}). The reason of using both types of variables is that the variation between subjects does not usually depend on the curve data only, summary statistics or some subject-specific variables do provide useful information, like the scenario in Figure \ref{fig:rec01} and Figure \ref{fig:rec02} in the simulated example and  Table \ref{tab:cluster_realdata} and Figure \ref{fig:real1} in the real data analysis. The introduction of the latent indicator variable is very useful in the implementation; see the details below.
	
	We call the models defined in (\ref{eq:mixt}) and (\ref{eq:allo}) as simultaneous registration and clustering (\textit{SRC}) models.
	
	\subsection{Estimation}
	\label{subsec:esti}
	The discrete form of model (\ref{eq:mixt}) for the $i$th curve data $\mf{x}_{ai} = \big(x_{ai}(t_{i1}),\dots, x_{ai}(t_{in_{i}})\big)^{\intercal}$ can be expressed as follows
	\begin{equation}\label{eq:mixt3}
	\ve{x}_{ai}|_{z_{ki} = 1} = \ve{\tau}_{ak}(g_{ki}) + \ve{r}_{aki} + \ve{\epsilon}_{i}, \hspace{5mm} \text{for}\hspace{2mm} a = 1, 2; \hspace{2mm} k = 1,\dots,K,
	\end{equation}
	where $\ve{\tau}_{ak}(g_{ki}) = \Big(\tau_{ak}\big(g_{ki}(t_{i1})\big), \dots, \tau_{ak}\big(g_{ki}(t_{in_{i}})\big)\Big)^{\intercal}$, and $\ve{r}_{aki}$ and $\ve{\epsilon}_{i}$  are both $n_{i}$-dimensional column vector. In this paper, we respectively set $\ve{S}$ as the Matern covariance function with parameters $\ve{\rho}_{s}$ and $\ve H$ as the unstructured covariance function or Brownian covariance function with parameter $\ve{\rho}_{h}$ \citep{Raket}, and they can be estimated by the data; the details are provided in the next subsection. Other covariance functions can also be used \citep{ShiB11}.  Let $\ve{S}_{aki}$ and $\ve{H}_{ki}$ be the covariance matrix of $\ve{r}_{aki}$ and $\ve{w}_{ki}$ respectively,  which can be calculated by the corresponding covariance function. We model $\tau_{ak}(t)$ using $q$ basis functions $\{\psi_{1}(t), \dots, \psi_{q}(t)\}$ with weights $\ve{d}_{ak} = (d_{ak1}, \dots, d_{akq})^\intercal$.
	Thus, $\ve{\tau}_{ak}(g_{ki}) = \ve{\Psi}_{ki}\ve{d}_{ak}$ where $\ve{\Psi}_{ki} = [\ve{\Psi}_{ki1}, \dots, \ve{\Psi}_{kiq}]_{n_{i} \times q}, \ve{\Psi}_{kil} = (\psi_{l}(g_{ki}(t_{i1})), \dots, \psi_{l}(g_{ki}(t_{in_{i}})))^{\intercal},  l=1,\dots,q$. In this paper, we use a smooth non-linear deformation for the curves, 
	which is produced by a cubic Hermite spline \citep{Raket}.
	
	The unknown parameters from the $k$-th component for the $a$-coordinates ($x$- or $y$-coordinates) of the $i$-th curve are denoted by $\ve{\theta}_{aki} \overset{\Delta}{=} \{\ve{d}_{ak},\ve{w}_{k}, \ve{w}_{ki}, \ve{\rho}_{s}, \ve{\rho}_{h}, \sigma\}$. Let $\ve{\theta}_{ki}$ be the vector of $\{\ve{\theta}_{aki}, a =1, 2\}$. We can similarly define $\ve{\theta}_{i} = \{\ve{\theta}_{ki}, k = 1, \dots, K\}$, $\ve{\theta} = \{\ve{\theta}_{i}, i = 1, \dots, N\}$ and $\ve{\beta}  = \{\ve{\beta}_{k}, k = 1, \dots, K-1\}$. 
	The Gaussian mixture distribution for the $i$-th curve data can be written in the form 
	\[
	p(\ve{x}_{i}|\ve{\theta}_{i},\ve{\beta})  = \sum_{k=1}^{K}\pi_{ki}p(\ve{x}_{i}|\ve{\theta}_{ki}),\hspace{5mm} i = 1, \dots, N,
	\]
	where $p(\ve{x}_{i}|\ve{\theta}_{ki}) = p(\ve{x}_{1i}|\ve{\theta}_{1ki})p(\ve{x}_{2i}|\ve{\theta}_{2ki})$. We assume $\ve{x}_{1i}$ and $\ve{x}_{2i}$ are conditional independent given those parameters. The log-likelihood of $(\ve{\theta}, \ve{\beta})$ is therefore
	\[
	L(\ve{\theta}, \ve{\beta}) = \sum_{i=1}^{N}\text{log}\Big\{\sum_{k=1}^{K}\pi_{ki}p(\ve{x}_{i}|\ve{\theta}_{ki})\Big\}.
	\]
	
	It is quite tricky to conduct the estimation due to the large number of unknown parameters. EM algorithm will be adopted in this paper. We have defined the latent indicator variable $\ve{z}_{i}$, which is treated as missing. The joint likelihood function of $\ve{x}$ and $\ve{z}$, where $\ve{z} = \{\ve{z}_{i}; i = 1, \dots, N\}$, takes the form
	\[
	p(\ve{x},\ve{z}|\ve{\theta}, \ve{\beta}) = p(\ve{z}|\ve{\beta})p(\ve{x}| \ve{z}, \ve{\theta}) = \prod_{i=1}^{N}\prod_{k=1}^{K}\pi_{ki}^{z_{ki}}p(\ve{x}_{i}|\ve{\theta}_{ki})^{z_{ki}}.
	\]
	Taking the logarithm, we have the log-likelihood for complete data $(\ve{x}, \ve{z})$
	\begin{align}
	L_{c}(\ve{\theta}, \ve{\beta}) = \text{log}p(\ve{x},\ve{z}|\ve{\theta}, \ve{\beta})  =  \sum_{i=1}^{N}\sum_{k=1}^{K}z_{ki}\bigg(\text{log}\pi_{ki} + \text{log}p(\ve{x}_{i}|\ve{\theta}_{ki})\bigg).
	\end{align}
	The expected value of the complete log-likelihood with respect to $\ve{z}$ is given by
	\begin{equation}{\label{eq:mle}}
	\begin{split}
	E_{\ve{z}}\{L_{c}(\ve{\theta}, \ve{\beta})\} & = \sum_{k=1}^{K}\sum_{i=1}^{N}E(z_{ki}|\ve{x}, \ve{\theta}, \ve{\beta})\bigg(\text{log}\pi_{ki} + \text{log}p(\ve{x}_{i}|\ve{\theta}_{ki})\bigg)\\
	& =  \sum_{k=1}^{K}\sum_{i=1}^{N}M_{ki}\bigg(\text{log}\pi_{ki} + \text{log}p(\ve{x}_{i}|\ve{\theta}_{ki})\bigg),
	\end{split}
	\end{equation}
	where
	\[
	M_{ki} \overset{\Delta}{=} E(z_{ki}|\ve{x}, \ve{\theta}, \ve{\beta}) = \frac{\pi_{ki}p(\ve{x}_{i}|\ve{\theta}_{ki})}{\sum_{j=1}^{K}\pi_{ji}p(\ve{x}_{i}|\ve{\theta}_{ji})},\hspace{5mm} i = 1, \dots, N; k = 1,\dots, K.
	\]
	The derivation of $M_{ki}$ is given by  \ref{ap:deri}. The procedure of EM algorithm includes
	\begin{enumerate}
		\item Initialize $\ve{\theta}^{(l)}$ and $\ve{\beta}^{(l)}$ and evaluate the $M_{ki}$ (E-step)
		\[
		M_{ki} = \frac{\pi_{ki}^{(l)}p(\ve{x}_{i}|\ve{\theta}^{(l)}_{ki})}{\sum_{j=1}^{K}\pi_{ji}^{(l)}p(\ve{x}_{i}|\ve{\theta}^{(l)}_{ji})} .
		\]
		
		\item Fix $M_{ki}$ and maximize $Q(\ve{\theta}, \ve{\beta})$ with respect to $\ve{\theta}$ and $\ve{\beta}$
		\[
		Q(\ve{\theta}, \ve{\beta})  \overset{\Delta}{=} \sum_{k=1}^{K}\sum_{i=1}^{N}M_{ki}\bigg(\text{log}\pi_{ki} + \text{log}p(\ve{x}_{i}|\ve{\theta}_{ki})\bigg),
		\]
		leading to $\ve{\theta}^{(l+1)}$ and $\ve{\beta}^{(l+1)}$ (M-step).
	\end{enumerate}
	The technical details are given in the next subsection.
	
	\subsection{Implementation}
	\label{subsec:impl}
	In  E-step, we first initialize the weights $M_{ki}$. In practice, we choose $M_{ki}^{(0)} \sim U(0,1)$ for the purpose of simplicity. Each $M_{ki}$ is then divided by their summation $\sum_{k=1}^{K}M_{ki}$ and we set $\pi^{(0)}_{ki} = M_{ki}^{(0)}$.
	In M-step, there are no analytic solutions to the maximization of $Q(\ve{\theta}, \ve{\beta})$ with respect to $\ve{\theta}$, so that we use the following algorithms. Maximizing $Q(\ve{\theta}, \ve{\beta})$ with respect to $\ve{\theta}$ given the current weights $M_{ik}$ is equivalent to maximizing
	\[
	\sum_{k=1}^{K}\sum_{i=1}^{N}M_{ki}\bigg(\sum_{a=1}^{2}\big(\text{log}p(\ve{x}_{ai}|\ve{\theta}_{aki})\big)\bigg).
	\]
	Borrowing the idea from \cite{Lars:2016}, all the parameters within $\ve{\theta}$ are estimated iteratively through three conditional models. In order to simplify the likelihood computations, all the random effects are scaled by a  noise standard deviation $\sigma$.  The norm induced by a full-rank covariance matrix $\ve{B}$ is denoted by $||\ve{A}||^{2}_{\ve{B}} = \ve{A}^{T}\ve{B}^{-1}\ve{A}$.

	\textbf{(i) Estimating the fixed effects $\ve{\tau}_{ak}$}
	
	Given $\ve{\theta}^{(l)}$, we have $\ve{x}_{ai}|_{z_{ki} = 1} \sim N_{n_{i}}(\ve{\Psi}_{ki}\ve{d}_{ak}, \ve{I}_{n_{i}}+\ve{S}_{aki}), a = 1, 2; i = 1, \dots, N$. $\ve{I}_{n_{i}}$ denotes the $n_{i} \times n_{i}$ identity matrix. So the negative log likelihood for the weights $\ve{d}_{ak}$ (its square magnitude is penalized by a weighting factor $\eta$) is proportional to
	\[
	L(\ve{d}_{ak}) = \sum_{i=1}^{N}M_{ki}||\ve{x}_{ai} - \ve{\Psi}_{ki}\ve{d}_{ak}||^{2}_{\ve{I}_{n_{i}}+\ve{S}_{aki}} + \eta\ve{d}_{ak}^\intercal\ve{d}_{ak}, \hspace{0.5cm} a = 1, 2; \hspace{0.5cm} k = 1, \dots, K.
	\]
	This gives the estimator
	\[
	\hat{\ve{d}}_{ak} =  (\ve{\Psi}_{k}^\intercal (\frac{\ve{I}_{n} + \ve{S}_{ak}}{\ve{M}_{k}})^{-1}\ve{\Psi}_{k} + \eta \ve{I}_{q})^{-1}\ve{\Psi}_{k}^\intercal (\frac{\ve{I}_{n} + \ve{S}_{ak}}{\ve{M}_{k}})^{-1}\ve{x}_{a}, \hspace{0.5cm} a = 1, 2; \hspace{0.5cm} k = 1, \dots,K,
	\]
	where $\ve{\Psi}_{k} = [\ve{\Psi}_{k1}^{\intercal},\dots,\ve{\Psi}_{kN}^{\intercal}]^\intercal \in \ve{R}^{m \times q}$, $m = \sum_{i = 1}^{N}n_{i}$, $\ve{x}_{a} = (\ve{x}_{a1}^{\intercal}, \dots, \ve{x}_{aN}^{\intercal})^{\intercal}$ and 
	\begin{equation}
	\frac{\ve{I}_{n}+\ve{S}_{ak}}{\ve{M}_{k}} \overset{\Delta}{=} \begin{bmatrix}
	(\ve{I}_{n_{1}}+\ve{S}_{ak1})/M_{k1} & &  \\
	& \ddots &  \\
	& & (\ve{I}_{n_{N}}+\ve{S}_{akN})/M_{kN}
	\end{bmatrix} \in \ms{R}^{m \times m}.
	\label{mk}
	\end{equation}
	
	\textbf{(ii) Estimating warping parameters $\ve{w}_{k}$ and $\ve{w}_{ki}$}
	
	Given $\ve{\theta}^{(l)}$ and $\hat{\ve{d}}_{ak}$, we have the joint probability density function of $(\ve{x}_{ai}, \ve{w}_{ki})$ given by
	\[
	p(\ve{x}_{ai}, \ve{w}_{ki}) = p(\ve{x}_{ai}|\ve{w}_{ki})*p(\ve{w}_{ki}) \sim N_{n_{i}}(\ve{\Psi}_{ki}\hat{\ve{d}}_{ak}, \ve{I}_{n_{i}}+\ve{S}_{aki}) * N_{n_{w}}(0, \ve{H}_{ki}).
	\]
	So, we can simultaneously estimate the fixed warping effects $\ve{w}_{k}$ and predict the random warping effects $\ve{w}_{ki}$ from the joint conditional negative log posterior. It is proportional to
	\begin{equation}\label{eq:post}
	L(\ve{w}_{k}, \ve{w}_{ki}) = \sum_{a=1}^{2}\sum_{i=1}^{N}M_{ki}||\ve{x}_{ai} - \ve{\Psi}_{ki}\hat{\ve{d}}_{ak}||^{2}_{\ve{I}_{n_{i}}+\ve{S}_{aki}} + 2\sum_{i=1}^{N}M_{ki}||\ve{w}_{ki}||^{2}_{\ve{H}_{ki}}, \hspace{0.3cm} k = 1, \dots, K,
	\end{equation}
	where $\ve{\Psi}_{ki}$ is determined by $n_{i}$ discrete values of the inverse of warping function $g_{ki}(t)$ which is totally characterized by $\ve{w}_{k}$ and $\ve{w}_{ki}$ as aforementioned.  By minimizing $L(\ve{w}_{k}, \ve{w}_{ki})$ we can obtain the estimation of $\ve{w}_{k}$ and the prediction of $\ve{w}_{ki}$.
	
	\textbf{(iii) Estimating the variance parameters $\sigma^{2}, \ve{\rho}_{s}$ and $\ve{\rho}_{h}$}
	
	By using the first-order Taylor approximation of model (\ref{eq:mixt3}) in the the random warping parameters $\ve{w}_{ki}$ around a given prediction $\ve{w}_{ki}^{0}$ ($\ve{w}^{0}_{ki}$ is specified by the estimate of $\ve{w}_{ki}$ from \textbf{(ii)} in the current iteration), we can write this model as a vectorized linear mixed-effects model
	\begin{equation}\label{eq:mod3}
	\ve{x}_{a}|_{z_{ki}=1} \approx \ve{G}_{ak} + \ve{B}_{ak}(\ve{W}_{k} - \ve{W}_{k}^{0}) + \ve{r}_{ak} + \ve{\epsilon},\hspace{0.5cm} a = 1, 2; \hspace{0.5cm} k = 1, \dots,K,
	\end{equation}
	where $\ve{x}_{a} = \{\ve{x}_{ai}, i = 1, \dots, N\}$,  with effects given by
	\begin{align*}
	\ve{G}_{ak} &= \Big\{\ve{\Psi}_{ki}|_{g_{ki} = g_{ki}^{0}}\ve{d}_{ak}\Big\}_{ij} \in \ms{R}^{m},\\
	\ve{B}_{ak}  &= \text{diag}(\ve{B}_{aki})_{i} \in \ms{R}^{m\times Nn_{w}},\\
	\ve{B}_{aki}  &=   \bigg\{\partial_{g_{ki}}\Big(\tau_{ak}\big(g_{ki}(t_{j})\big)\Big)\Big|_{g_{ki} = g_{ki}^{0}}\Big(\nabla_{\ve{w}_{ki}}\big(g_{ki}(t_{j})\big)\Big)^\intercal\Big|_{\ve{w}_{ki} = \ve{w}_{ki}^{0}}\bigg\}_{j} \in \ms{R}^{n_{i} \times n_{w}},\\
	\ve{W}_{k}  &=  (\ve{w}_{ki})_{i} \sim N_{Nn_{w}}(0, \sigma^{2}\ve{I}_{N}\otimes \ve{H}_{n_{w} \times n_{w}}),  \hspace{3mm} \ve{W}^{0}_{k}  =  (\ve{w}^{0}_{ki})_{i} \in \ms{R}^{Nn_{w}},\\
	\ve{r}_{ak} &\sim N_{m}(0,\sigma^{2}\ve{S}_{ak}), \hspace{3mm} \ve{S}_{ak}  = \text{diag}(\ve{S}_{aki})_{i} \in \ms{R}^{m \times m},\\
	\ve{\epsilon} &\sim N_{m}(0,\sigma^{2}\ve{I}_{m}),
	\end{align*}
	where $g^{0}_{ki}(t) = t + w_{k}(t) + w^{0}_{ki}(t)$. $\text{diag}(\ve{B}_{aki})_{ki}$ is the block diagonal matrix with the $\ve{B}_{aki}$ matrices along its diagonal, so is $\text{diag}(\ve{S}_{aki})_{i}$.  The derivation of the linearized model (\ref{eq:mod3}) is given in  \ref{ap:linear}.
	The negative profile log likelihood function for the model (\ref{eq:mod3}) is proportional to
	\[
	L(\sigma^{2}, \ve{\rho}_{s}, \ve{\rho}_{h}) = \sum_{k=1}^{K}\Big\{\sum_{a = 1}^{2}\sigma^{2}||\ve{x}_{a} - \ve{G}_{ak} + \ve{B}_{ak}\ve{W}_{k}^{0}||^{2}_{\ve{V}_{ak}} + \sum_{a=1}^{2}\text{log det}\ve{V}_{ak}\Big\} + 2mK\text{log} \sigma^{2},
	\]
	where $\ve{V}_{ak} = \frac{\Big(\ve{S}_{ak} + \ve{B}_{ak}\big(\ve{I}_{N} \otimes\ve{H}_{n_{w}\times n_{w}}\big)\ve{B}_{ak}^\intercal + \ve{I}_{m}\Big)}{\ve{M}_{k}} \in \ms{R}^{m \times m}$, with the definition similar to $\frac{\ve{I}_{n}+\ve{S}_{ak}}{\ve{M}_{k}}$ in \eqref{mk}.
	
	To speed up convergence, we usually repeat the above three steps several times within each iteration. 
	
	\textbf{(iv) Updating $M_{ki}$ and estimating $\bs{\beta}$}
	
	Fix $\ve{\theta}^{(l+1)}$ and update
	\[
	M_{ki} = \frac{\pi^{(l)}_{ki}p(\ve{x}_{i}|\ve{\theta}^{(l+1)}_{ki})}{\sum_{j=1}^{K}\pi^{(l)}_{ji}p(\ve{x}_{i}|\ve{\theta}^{(l+1)}_{ji})},
	\]
	where
	\[
	\pi_{ki}^{(l)} = \frac{\text{exp}\{\ve{v}^{\intercal}_{i}\ve{\beta}_{k}^{(l)}\}}{1 + \sum_{j=1}^{K-1}\text{exp}\{\ve{v}^{\intercal}_{i}\ve{\beta}_{j}^{(l)}\}}, \hspace{5mm} k = 1,\dots, K-1,
	\]
	and $\pi_{Ki} = 1 - \sum_{j=1}^{K-1}\pi_{ji}^{(l)}$. Then maximize $Q(\ve{\theta}, \ve{\beta})$ with respect to $\ve{\beta}$, which is equivalent to maximize
	\[
	L(\ve{\beta}) \overset{\Delta}{=} \sum_{i = 1}^{N}\Bigg\{\sum_{k=1}^{K-1}M_{ki}\bigg\{\ve{v}_{i}^{\intercal}\ve{\beta}_{k} - \text{log}\big[1 + \sum_{j = 1}^{K - 1}\text{exp}\{\ve{v}_{i}^{\intercal}\ve{\beta}_{j}\}\big]\bigg\} - M_{Ki}\text{log}\big[1 + \sum_{j = 1}^{K - 1}\text{exp}\{\ve{v}_{i}^{\intercal}\ve{\beta}_{j}\}\big]\Bigg\}.
	\]
	This is very similar to the log-likelihood for a multinomial logit model ($M_{ki}$'s are corresponding to the observations) and can be maximized by iteratively re-weighted least square algorithm.
	
	\subsection{Model selection, clustering and related methods}
	\label{subsec:mod_sel}
	There are two questions on the model selection for our proposed simultaneous registration and clustering method  for multi-dimensional functional data: one is how to determine the number of knots for the splines and another is how many clusters. For the former, since our data is rather dense and insensitive,  it works well using a relatively small number of equally-spaced knots. For the choice of the number of clusters, $K$, since the number of parameters, $P_{L}$, in model (\ref{eq:mixt}) is  relative to the number of subjects, $N$, a second-order bias correction version  of AIC called $\text{AIC}_{c}$ \citep{Sug78, Ken04} is utilized:
	\[
	\text{AIC}_{c} = -2L(\bs{\hat{\Theta}}) + 2P_{L} + \frac{2P_{L}(P_{L} + 1)}{N - P_{L} - 1},
	\]
	where $L(\ve{\hat{\Theta}})$ is the maximized log-likelihood function, $\ve{\hat{\Theta}} = \{\ve{\hat{\theta}}, \hat{\ve{\beta}}\}$ in this paper.
	
	In inference,  we first choose $K$ clusters by $\text{AIC}_{c}$. Then fit the data using the method discussed in the previous subsections and denote the estimates of the parameters by $\hat{\ve{\theta}}$ and $\hat{\ve{\beta}}$. Under the framework of SRC method, the fixed-effect part of the $i$th individual curve is calculated by
	\begin{equation}\label{eq:pred}
	\hat{\ve{x}}_{ai}(t) = \sum_{k=1}^{K}\hat{\pi}_{ki}\big[\hat{\tau}_{ak}(\hat{g}_{ki}(t))\big], \hspace{2mm}a = 1, 2; \hspace{2mm} i = 1,\dots, N,
	\end{equation}
	where $\hat{g}_{ki}(t) = t + \hat{\ve{w}}_{k}(t) + \hat{\ve{w}}_{ki}(t)$ and $\hat{\pi}_{ki} = \frac{\text{exp}\{\ve{v}_{i}\hat{\ve{\beta}}_{k}\}}{1 + \sum_{j=1}^{K-1}\text{exp}\{\ve{v}_{i}\hat{\ve{\beta}}_{j}\}}$.
	
	For any individual data $D^{\ast} = \{(\ve{x}_{1\ast},\ve{x}_{2\ast}), \ve{v}^{\ast}\}$ in $D$, the posterior distribution of the cluster membership $\ve{z}^{\ast} = (z_{1}^{\ast}, \dots, z_{K}^{\ast})^{\intercal}$ is given by
	\[
	p(z_{k}^{\ast} = 1|D^{\ast}) = \frac{\pi_{k}^{\ast}p(\ve{x}_{1*}|\hat{\ve{\theta}}_{1k})p(\ve{x}_{2*}|\hat{\ve{\theta}}_{2k})}{\sum_{j=1}^{K_{0}}\pi_{j}^{\ast}p(\ve{x}_{1*}|\hat{\ve{\theta}}_{1j})p(\ve{x}_{2*}|\hat{\ve{\theta}}_{2j})},
	\]
	where
	\[
	\pi_{k}^{\ast} = \frac{\text{exp}\{\ve{v}^{\ast \intercal}\hat{\ve{\beta}}_{k}\}}{1 + \sum_{j=1}^{K-1}\text{exp}\{\ve{v}^{\ast \intercal}\hat{\ve{\beta}}_{j}\}}.
	\]
	As a result, the best cluster membership for $D^{\ast}$ can be determined by
	\[
	k^{\ast}  = {\argmax}_{k = 1,\dots,K}\{p(z_{k}^{\ast} = 1|D^{\ast})\}.
	\]
	The average mean curve for each group can be calculated from $\{\hat{\ve{x}}_{ai}(t)|_{z_{ki} = 1}\}$ for $k=1, \ldots, K$. 
	
	\subsubsection{Related methods}
	\label{subsubsec:rela}
	Functional $k$-means method is a popular approach for clustering curves \citep{Chiou07}, which is an extension of $k$-means cluster \citep{Mac67, Lloyd82} for scalar variables. The idea can be extended to do clustering and registration simultaneously. Using the similar notation around (\ref{eq:mixt3}), we can define the following
	objective function
	\begin{equation}\label{eq:kmean}
	F = \sum_{i=1}^{N}\sum_{k=1}^{K}z_{ki}d\big(\ve{x}_{i}, \ve{\tau}_{k}(g_{ki})\big),
	\end{equation}
	where $d$ represents one kind of distance between each curve to its assigned mean curve $\ve{\tau}_{k}(g_{ki})$ and
	\[
	z_{ki}  = \left\{
	\begin{array}{ll}
	1, \hspace{2mm}\text{if} \hspace{2mm}k = \argmin_{j}d\big(\ve{x}_{i}, \ve{\tau}_{j}(g_{ji})\big),\\
	0, \hspace{2mm}\text{otherwise}.
	\end{array}
	\right.
	\]In order to find the values $\{z_{ki}\}$ and the $\{\ve{\tau}_{k}(g_{ki})\}$ to minimize $F$, we  can perform an iterative procedure in which each iteration involves two steps of the optimization with respect to $\{z_{ki}\}$ and  $\{\bs{\tau}_{k}(g_{ki})\}$ respectively; the details are given in  \ref{ap:imp}. This approach is denoted  by \textit{k-means-f}. 
	
	A special case of  the SRC model  defined in Section \ref{subsec:mode} is that the allocation model in (\ref{eq:allo}) doesn't depend on any scalar variables (denoted by \textit{SRC-f}, i.e. use the function variable only). This special case is very similar to the above \textit{k-means-f} approach.
	Actually the \textit{k-means-f}  algorithm is a special case of EM algorithm for \textit{SRC-f}. Using the similar notation around (\ref{eq:mixt3}) and assuming
	$
	\ve{x}_{ai}|_{z_{ki} = 1} \sim N_{n_{i}}(\ve{\tau}_{ak}(g_{ki}), \delta\ve{I}), a = 1, 2; i = 1, \dots, N,
	$
	where $\delta$ is shared by all the clusters, we have the density function of $\ve{x}_{ai}$ with the form
	\[
	p(\ve{x}_{ai}|\ve{\theta}_{aki}) = (2\pi\delta)^{-\frac{n_{i}}{2}}\text{exp}\big\{-\frac{1}{2\delta}||\ve{x}_{ai} - \ve{\tau}_{ak}(g_{ki})||^{2}\big\}.
	\]
	Let the allocation model as $p(z_{ki} = 1) = \pi_{k}, k = 1, \dots, K$ with $\sum_{k=1}^{K}\pi_{k} = 1$. Using the EM algorithm for the Gaussian mixtures described in Section \ref{subsec:esti}, we have
	\[
	M_{ki} = \frac{\pi_{k}\prod_{a=1}^{2}\text{exp}\big\{-||\ve{x}_{ai} - \ve{\tau}_{ak}(g_{ki})||^{2}/2\delta\big\}}{\sum_{j=1}^{K}\pi_{j}\prod_{a=1}^{2}\text{exp}\big\{-||\ve{x}_{ai} - \ve{\tau}_{aj}(g_{ji})||^{2}/2\delta\big\}}.
	\]
	Clearly, $M_{ki} \rightarrow z_{ki}$, when $\delta \rightarrow 0$. Thus $E_{\ve{z}}[\text{log}p(\ve{x},\ve{z}|\ve{\theta}, \ve{\beta})] \approx -\frac{1}{2\delta}F + \text{costant}$, when $\delta$ is small.
	It means the optimization problem is the same as the \textit{k-means-f} algorithm given by (\ref{eq:kmean}) (use $d = ||\cdot||^{2}$).
	
	\section{Numerical analysis}
	\label{sec:num_ana}
	We shall evaluate the performance and properties of the proposed \textit{SRC} model in this section. We will compare it with functional $k$-means clustering (\textit{k-means-f}) with simultaneous registration as discussed in Section \ref{subsubsec:rela}, the \textit{SRC} without using an allocation model (\textit{SRC-f}) and scalar $k$-means clustering (\textit{k-means-s}). The \textit{k-means-s} is a general k-means clustering method using scalar variables only. We will conduct  analysis on both simulated and real data.
	
	\subsection{Simulation study} \label{sec:simu}
	In this simulation study, we consider 2D curves coming from two groups. 
	For each group, the corresponding observations of functional variables $\ve{x}(t)$ and scalar variables $v$ will be generated. There are $N_{k}$ batches of data in each group, where $k$ = 1 and 2. We will evaluate and compare four methods based on the simulated data $D = \{(\ve{x}_{i}(t_{ij}), \ve{v}_{i}); i = 1,\dots, N; j = 1,\dots, n_{i}\}$ in different scenarios where $N=N_{1}+N_{2}$.
	
	\subsubsection{Data generation}
	\begin{enumerate}
		\item Generate the underlying true curves, i.e. the curves  based on the internal time scale $g^{-1}(t)$. We first assume those curves in two groups share the following two slightly different true means, which is  similar to two patterns of  those real curves of hyoid bone's movement
		\begin{equation}
		\begin{split}
		\bs{\mu}_{1}(t) & = \big(\mu_{11}(t), \mu_{21}(t)\big) = \big(\text{exp}\{\text{cos}(2\pi t)\}, \text{exp}\{\text{sin}(2\pi t)\}\big), \\
		\bs{\mu}_{2}(t) & =  \big(\mu_{12}(t), \mu_{22}(t)\big) = \big(\text{exp}\{\text{cos}(2\pi t^{1.05} - b_{1})\}, \text{exp}\{\text{sin}(2\pi t^{1.1} + b_{1})\}\big).
		\end{split}
		\end{equation}
		
		The degree of overlapping between two groups relies on the value of $b_{1}$. The smaller the value of $b_{1}$, the higher the degree of overlapping, and more difficult to cluster those curves. We use the equidistant points $t_{j} = \frac{j+1}{102}, j = 1,\dots, 100$ as input grid, i.e. $n_i=100$. The underlying true curves are generated as
		\[
		\ve{x}_{ak}(g^{-1}(t)) = \mu_{ak}(t), \hspace{5mm} a = 1, 2; \hspace{2mm} k = 1, 2.
		\]
		Figure \ref{fig:cluster_true_mean_d010} in  \ref{ap:sist} shows the shape of the true mean curves for different values of $b_{1}$.
		
		\item Generate the original 2D curves $\ve{x}(t)$ by adding the warping function, amplitude variation and errors as follows.
		\begin{enumerate}
			\item Model the true curves $\ve{x}_{ak}(g^{-1}(t))$  using B-spline basis function with 8 knots and obtain the coefficients $\ve{d}_{ak}$.
			\item For simplicity, we set $g_{ki}(t) = t +  w_{ki}(t)$ and use hyman spline (monotone cubic spine using Hyman filtering) based on the anchor knots $t_{w} = (0, 0.33, 0.67, 1)$ ($n_{w} = 4$). Set $\ve{w}_{ki} \sim N_{4}(0, \ve{T}_{k}^{\intercal}\ve{\Gamma}_{i})$, where $\ve{T}_{k}^{\intercal}\ve{T}_{k} = \ve{O}_{k}$,  $\ve{O}_{1} = \begin{bmatrix} 10 & 4 \\ 4 & 8\end{bmatrix}$ and $\ve{O}_{2} = \begin{bmatrix} 10 & 8 \\ 8 & 15\end{bmatrix}$,  and $\ve{\Gamma}_{i} = (\ve{\Gamma}_{i1}, \ve{\Gamma}_{i2})^{\intercal}$ with $\ve{\Gamma}_{i1}, \ve{\Gamma}_{i2}$ being independent random variables $N(0, \sigma^{2}_{w})$, where $i = 1, \dots, N_{k}$.
			\item Set the amplitude variation $\ve{r}_{aki} = \ve{T}_{0}^{\intercal}\cdot\ve{\Gamma}_{i0}$, where $\ve{T}_{0}^{\intercal}\ve{T}_{0} = \ve{O}_{0}$, $a = 1, 2$, $k = 1, 2$, $i =1, \dots, N_{k}$. $\ve{O}_{0}$ is created by Matern covariance function with $\ve{\rho}_{r} = (100, 0.3, 3)$, where the three elements represent the scale, range and smoothness,  respectively \citep{Raket},  and $\bs{\Gamma}_{j0}$ is a vector of 100 independent  normal random variables $N(0, \sigma^{2}_{r})$. Set $\bs{\epsilon} \sim N(0, \sigma^{2}\ms{I})$. 
			\item Generate $\ve{x}(t)$ based on the model (\ref{eq:mixt3}).
		\end{enumerate}
		
		\item Generate $\ve{v}$'s. We next generate those scalar variables $v$ by sampling from uniform distribution as follows:
		\[
		V_{i}  \sim \left\{
		\begin{array}{ll}
		\text{U}(1, 2), \hspace{2mm} i = 1, 2, \dots, N_{1},\\
		\text{U}(1 - b_{2}, 2 - b_{2}), \hspace{2mm} i = N_{1}+1,  \dots, N_{1}+N_{2}.
		\end{array}
		\right.
		\]
		Bear in mind that the larger the value of $d_{2}$, the lower the degree of overlapping and easier to carry out clustering using scalar variables.
	\end{enumerate}
	
	\subsubsection{Results}
	\label{subsec:res}
	In order to investigate how the overlapping of the observations of both  scalar variables and functional variables affect the performance of clustering, we study four scenarios with the constraint $4\sigma_{w}^{2} = \sigma_{r}^{2} = \sigma^{2} = 0.01^2$ and $N_{1} = N_{2} = 30$, and with 100 replications for each scenario. We use two measures to assess the performance of each method. The first one is the Rand index (RI) \citep{Ran:71}, having a value between 0 and 1, with 0 indicating two data clusterings disagree on any pair while 1 indicating a perfect match. And the second one is called adjusted Rand index (ARI) \citep{Hub:85}, a modified version of Rand index (ARI). A larger value of RI or ARI indicates a higher agreement of the method and the truth.
	
\begin{figure}	
		\centering
		\begin{subfigure}[t]{0.4\textwidth}
			\centering
			\includegraphics[width= \textwidth]{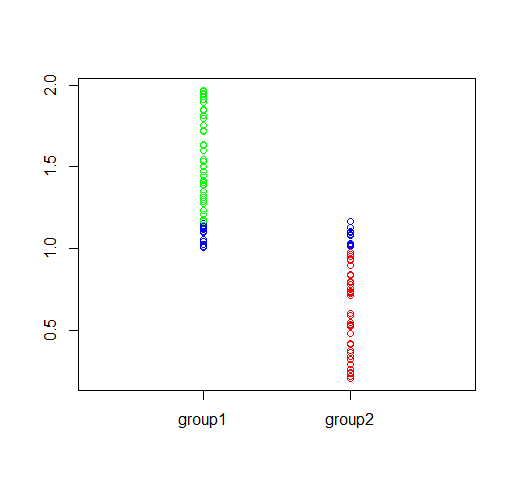}
			\caption{$b_{2} = 0.8$}		
		\end{subfigure}
		\quad
		\begin{subfigure}[t]{0.4\textwidth}
			\centering
			\includegraphics[width=\textwidth]{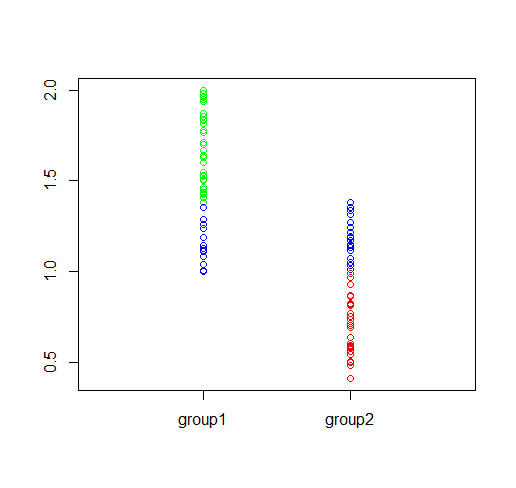}
			\caption{$b_{2} = 0.6$}
		\end{subfigure}
		\caption{Observations of scalar variable in two cases. The `blue' ones stand for those in the range of overlapping.}\label{fig:scal}
	\end{figure}

\begin{figure}	
		\centering
		\begin{subfigure}[t]{\textwidth}
			\centering
			\includegraphics[width= 0.3\textwidth]{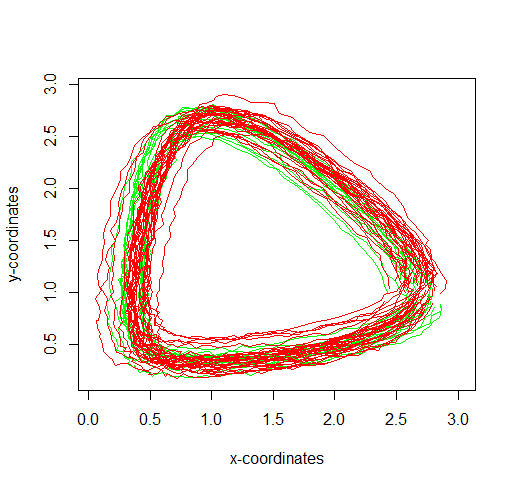}
			\includegraphics[width=0.3\textwidth]{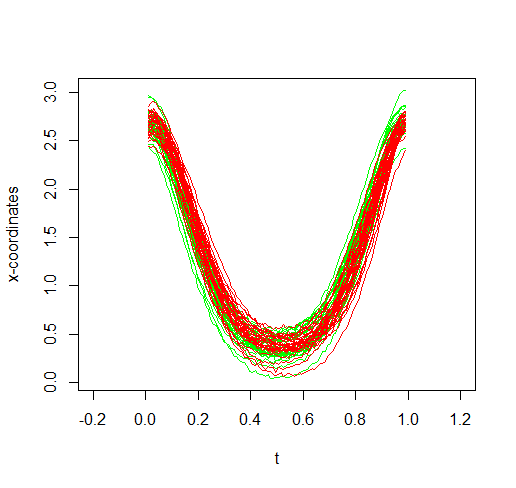}
			\includegraphics[width=0.3\textwidth]{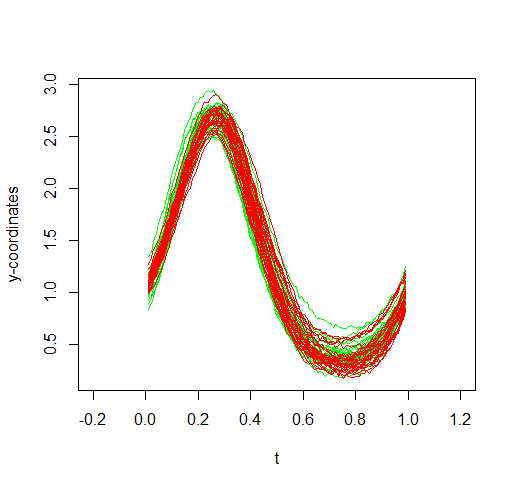}
			\caption{$b_{1} = 0.08$}		
		\end{subfigure}
		\quad
		\begin{subfigure}[t]{\textwidth}
			\centering
			\includegraphics[width= .3\textwidth]{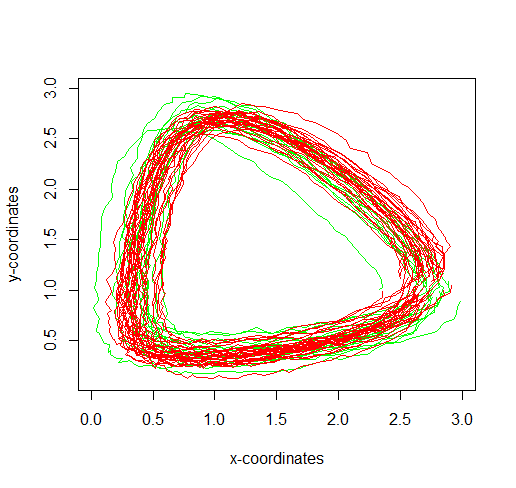}
			\includegraphics[width=.3\textwidth]{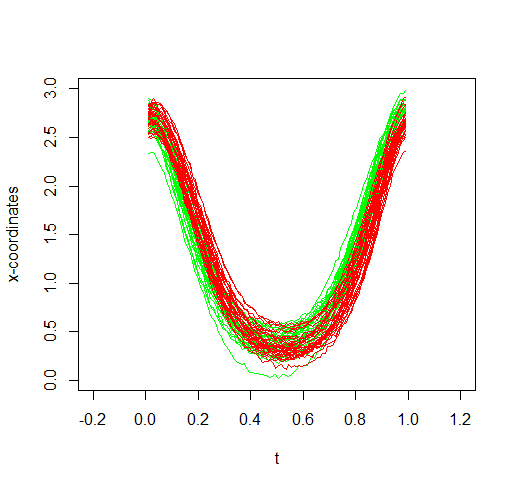}
			\includegraphics[width=.3\textwidth]{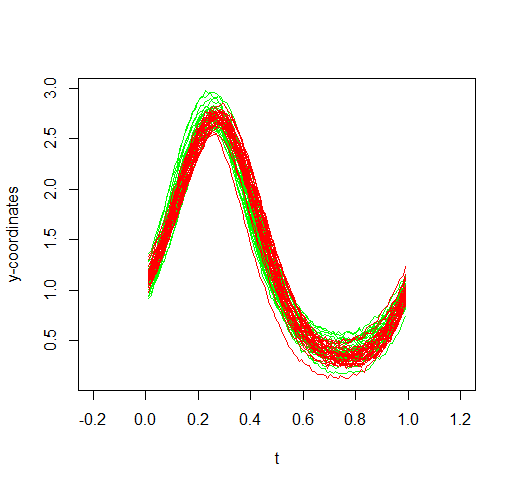}
			\caption{$b_{1} = 0.10$}		
		\end{subfigure}
		\quad
		\begin{subfigure}[t]{\textwidth}
			\centering
			\includegraphics[width= .3\textwidth]{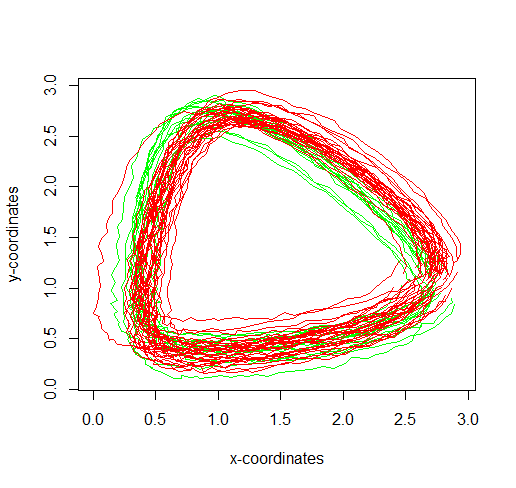}
			\includegraphics[width=.3\textwidth]{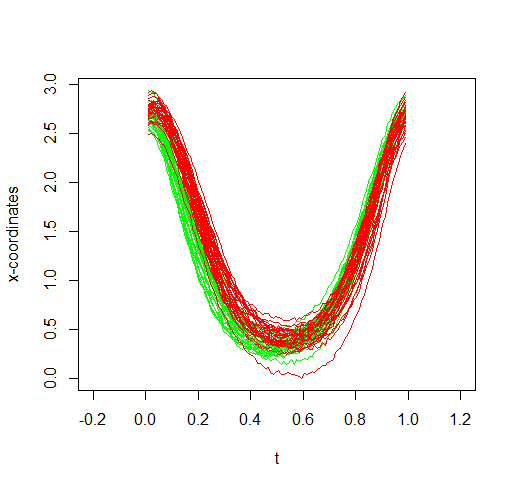}
			\includegraphics[width=.3\textwidth]{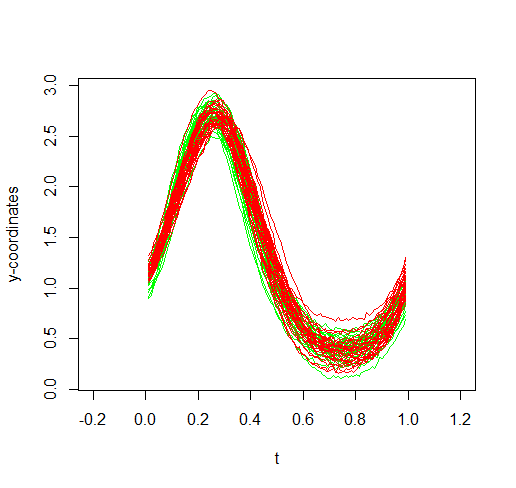}
			\caption{$b_{1} = 0.12$}		
		\end{subfigure}
		\caption{The raw 2D curves in one simulation run in three cases. }\label{fig:raw}
	\end{figure}

	Four methods are applied to the simulated data $D$ in Scenario 1 with $b_{1} = 0.12$, $b_{2} = 0.8$, Scenario 2 with $b_{1} = 0.10$, $b_{2} = 0.8$, Scenario 3 with $b_{1} = 0.08$, $b_{2} = 0.8$ and Scenario 4 with $b_{1} = 0.08$, $b_{2} = 0.6$. Figure \ref{fig:scal} and Figure \ref{fig:raw} show the raw data depending on different $b_{2}$ and $b_{1}$ respectively. First of all, we apply the $\text{AIC}_{c}$ to choose the number of clusters. The results from Figure \ref{fig:AICc} in  \ref{ap:sist} show that $\text{AIC}_{c}$ score reaches its minimum at 2 clusters. Table \ref{tab:clust} summarizes the comparisons by average ARI and RI. Overall, both measures suggest that the proposed \textit{SRC} outperform the other three methods in all scenarios because of the use of  both functional and scalar data. 
	
	From Table \ref{tab:clust},  we note that all the four methods perform best in Scenario 1 compared to the other scenarios. In this scenario, both $b_{1}$ and $b_{2}$ take the largest values, indicating that overlapping of the functional data (Figure \ref{fig:raw} (c)) and scalar data (Figure \ref{fig:scal} (a)) are the smallest and both greatly contribute to distinguishing those two clusters. Though the other three methods,  \textit{SRC-f}, \textit{k-means-f} and \textit{k-means-s} based on either functional data or scalar data,  also have good performance but not as good as \textit{SRC}.
	
	The first three scenarios share the same value of $b_{2}$, indicating that the degree of overlapping in two clusters for scalar data does not change (Figure \ref{fig:scal} (a)). It causes the performance of \textit{k-means-s} remaining the same. The overlapping in two clusters for functional data,  however,  gets smaller and smaller as the value of $b_{1}$ decreases from Scenario 1 to Scenario 3. It leads to a sharp decline for the performance of \textit{SRC-f} and \textit{k-means--f}, both of which  depend on functional data only, as opposed to a mild decrease of the performance of \textit{SRC}, which is based on both scalar data and functional data. 
	
	The scenario 4  has the smallest $b_{1}$ (Figure \ref{fig:raw} (a)) and $b_{2}$ (Figure \ref{fig:scal} (b)) and it is quite difficult to carry out clustering just based on functional data or scalar data only. Consequently, the values of ARI for \textit{SRC-f}, \textit{k-means-f} and \textit{k-means-s} are very small. But \textit{SRC} still performs well and are much better than the others. 
	
	Other combinations with varying overlapping determined by $b_{1}$ and $b_{2}$ and with different sample sizes have also been examined. The results presented here are very typical.
	
	\begin{table}[ht!]
		\centering
		\begin{tabular}{cccccccccccccc}
			\hline
			&&&\multicolumn{2}{c}{Scenario 1}&&\multicolumn{2}{c}{Scenario 2}&&\multicolumn{2}{c}{Scenario 3}&&\multicolumn{2}{c}{Scenario 4}\\
			\cline{4-5}\cline{7-8}\cline{10-11}\cline{13-14}
			&&&RI&ARI&&RI&ARI&&RI&ARI&&RI&ARI\\
			\hlineB{4}
			\textit{SRC}                                                               &&&$\mf{0.98}$&$\mf{0.96}$&                &$\mf{0.95}$&$\mf{0.90}$                  &&$\mf{0.91}$&$\mf{0.82}$                 &&$\mf{0.80}$&$\mf{0.61}$\\
			\textit{SRC-f}                                                                   &&&0.90&0.80                                            &&0.75&0.49                                         &&0.62&0.25                         &&0.62&0.25\\
			\textit{k-means-f}                                                                &&&0.91&0.83                                            &&0.76&0.53                                         &&0.64&0.29                         &&0.64&0.29\\
			\textit{k-means-s}                                                                     &&&0.81&0.62                                            &&0.81&0.62                                         &&0.81&0.62                          &&0.69&0.37\\
			\hline
		\end{tabular}
		\caption{Comparison of average clustering results among four methods.} \label{tab:clust}
	\end{table}
	
	\subsubsection{Recovery of curves and cluster patterns}
	To understand the underlying process better, it is necessary to use the optimally alignment to estimate the entire curve, so we estimate the aligned individual curves and reconstruct the cluster pattern using equation (\ref{eq:pred}).
	
	Figure \ref{fig:rec1} in Appendix displays one simulation run of $N = 100$, with $N_{1} = N_{2} = 50$ curves  in each group. The top panel presents the original raw curves in the two clusters in two colors in two dimensions ($x$-axis and $y$-axis) for a new scenario with $4\sigma_{w}^{2} = \sigma_{r}^{2} = \sigma^{2} = 0.02^2, b_{1} = 0.15, b_{2} = 0.8$. The other panels respectively show the individual aligned curves resulting from \textit{SRC}, \textit{SRC-f} and \textit{k-means-f}, with the value of RI (1, 0.63, 0.79) and the value of ARI (1, 0.26, 0.54) respectively.  The \textit{SRC} properly differentiates the two clusters (red and green) after curve alignment and performs better in recovering the cluster patterns.
	
\begin{figure}	
		\centering
		\begin{subfigure}[t]{0.45\textwidth}
			\centering
			\includegraphics[width= \textwidth]{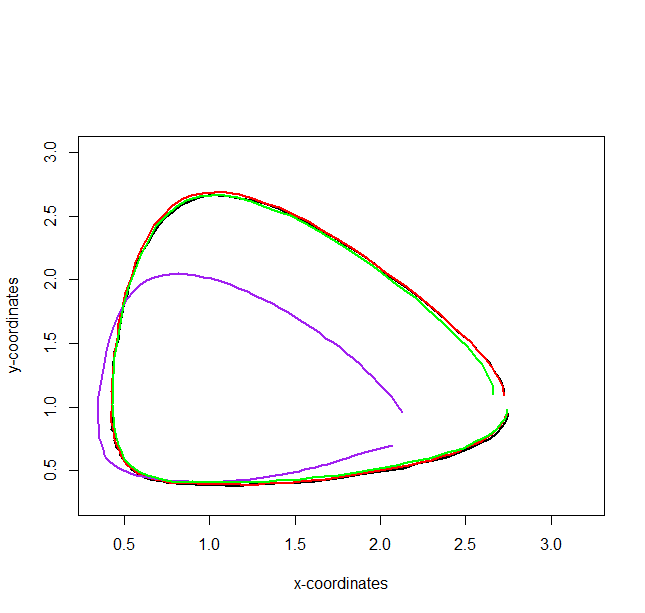}
			\caption{Cluster 1}		
		\end{subfigure}
		\quad
		\begin{subfigure}[t]{0.45\textwidth}
			\centering
			\includegraphics[width= \textwidth]{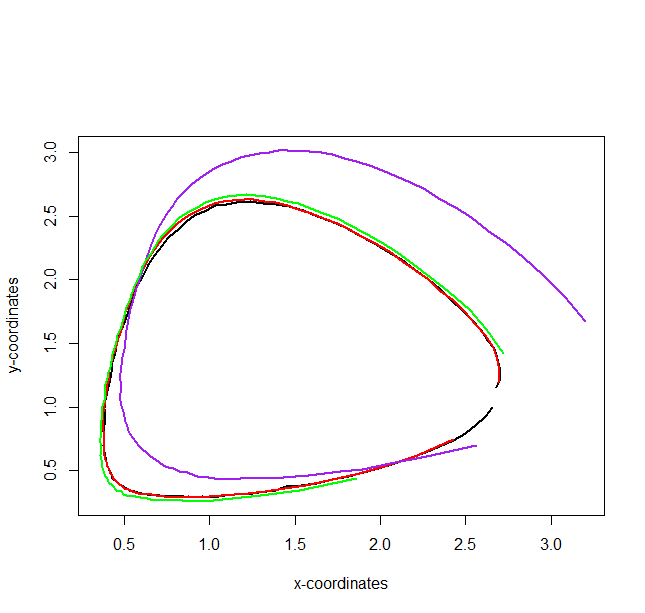}
			\caption{Cluster 2}		
		\end{subfigure}	
		\caption{Mean functions for 2D curves in each cluster. \emph{black lines} are true mean curves. \emph{red lines}, \emph{purple lines} and \emph{green lines} stand for mean curves calculated from the results from \textit{SRC}, \textit{SRC-f} and \textit{k-means-f} respectively.}\label{fig:rec2}
	\end{figure}

	Figure \ref{fig:rec2} summarizes the result of clustering patterns. 
	It shows  the SRC method recovered the true pattens very well. As a measure of estimation error, we use the root average squared error \citep{Ger04}
	\[
	\textit{rase}(\hat{\ve{\mu}}) = \sqrt{\frac{\sum_{j=1}^{m}||\hat{\ve{\mu}}(t_{j}) - \ve{\mu}(t_{j})||^2}{m}},
	\]
	where $m$ is the number of observation points,  $\ve{\mu}(t)$ and $\hat{\ve{\mu}}(t)$ are the cross-sectional mean of 2D raw curves and of registered curves from the target method respectively. The values of \textit{rase} are 2.9, 4.6 and 5.4  corresponding to three models \textit{SRC}, \textit{SRC-f} and \textit{k-means-f}.
	
	One extreme scenario has been discussed in \ref{ap:exts}, providing further evidence of the good performance of the proposed \textit{SRC}.
	
	\subsection{Real data analysis}
	\label{subsec:real}
	The application to a real data is to cluster the normal people and the patients with stroke  by studying their hyoid bone motion as well as the other scalar variables. Two groups, one for normal people and the other for patients,  are included.
	Figure (\ref{fig:realdata})(a) shows one frame from a X-ray video clip. The position of hyoid bone is tracked in each frame by a semi-automatic programme developed in \cite{Kim17}. The raw data    before being preprocessed are shown in Figure (\ref{fig:realdata})(b). There are  15 subjects in each group. The scalar variable we choose is the size of Pyriform Sinus Residue (see its position in Figure \ref{fig:pyriform}). Regarding to those 2D curves, we firstly carry out the preprocessing procedures like multi-dimensional shift, scaling and rotation using the package of GPA. While modeling those curves, we assume the warping function be a smooth nonlinear deformation  produced by an increasing spline and the random vector $\bs{w}_{ki}$ be a Brownian bridge observed at discrete anchor points. B-spline basis functions are utilized for modeling the mean curves. The  covariance function for the amplitude variance is the Matern covariance function. 
	
	We examine the performance of four methods \textit{SRC}, \textit{SRC-f}, \textit{k-means-f} and \textit{k-means-s} aforementioned. The values of AICc are shown in Figure \ref{fig:AICc_real1}. It shows that the two-component mixture model has the smallest  value. Table \ref{tab:cluster_realdata} shows the values of RI and ARI by comparing clustering results by the four methods with the clinic outcomes.  We can see that the \textit{SRC} method outperforms the other three. As a matter of fact,  both \textit{SRC-f} and \textit{k-means-f} with value of RI equivalent to 0.5 fail in this real data example. It is similar to the extreme example in  \ref{ap:exts}. More numerical results are provided in  \ref{sec:appreal}. 
	
	\begin{table}[ht!]
		\centering
		\begin{tabular}{c|cc}
			\hline
			Model  & RI & ARI\\
			\hline
			\textit{SRC}&$\bs{0.71}$&$\bs{0.42}$\\
			\textit{SRC-f}&0.50&0.02\\
			\textit{k-means-f}&0.50&0.02\\
			\textit{k-means-s}&0.67&0.33\\
			\hline
		\end{tabular}
		\caption{Results of clustering by four methods for the real data}\label{tab:cluster_realdata}
	\end{table}
	
	\section{Conclusion}
	\label{sec:dis}
	
	We have proposed a methodology for simultaneous registration and clustering, SRC,  for multi-dimensional functional data which considers both the curves and scalar variables. This model captures the heterogeneity from the potential time warping for curves and scalar variables corresponding to each subject while carrying out the clustering in the  meantime. It can be implemented with EM algorithm. Numerical examples show that it outperforms three other related methods, \textit{SRC-f}, \textit{k-means-f} and \textit{k-means-s}. The results in Section \ref{subsec:res} show that in most cases the inclusion of scalar variables can improve the performance of clustering  in functional data analysis.
	
	Generally, the registration for multi-dimensional functional data is much more complicated than one dimensional case. We use the pre-processing package  GPA \citep{Gower:1975} and a further registration via a simple warping function. The latter is  one of the key parts in our model. This approach performs very well in the numerical examples presented in this article. Further research is however needed, for example, how to improve the iterative implementation for the complete registration, similar to the shape geodesic algorithm by the metric-based method proposed by \cite{Sri11a}. The inverse of warping function $g$ in model (\ref{eq:mixt}) can also be replaced with various types of other functions depending on types of data. The success of resolving registration problem often depends on the flexibility of choosing warping function. 
	
	Model selection is an interesting but difficult issue for a mixture model, especially for the models with complex forms. \cite{Ken04} suggested AICc should be used unless $\frac{N}{P_{L}}>40$ for the model with the large value of $P_{L}$. In our model, the number of parameters $P_{L}$ is quite close to the number of subject $N$. Thus, we use AICc. It works well for the examples discussed in the article.  It is worth a further study under a general functional data analysis framework. 
	
	The results we obtained for the real data are encouraging although it is still in early stage. Research for this topic is carrying on. More features extracted from video clips along with other variables, both functional and scalar,  are under investigation. Different types of models for data fitting, clustering/classification and prediction are being developed.   
	

\begin{thebibliography}{}
	
	\bibitem[Bouveyron and Jacques, 2011]{Bou11}
	Bouveyron, C. and Jacques, J. (2011).
	\newblock Model-based clustering of time series in group-specific functional
	subspaces.
	\newblock {\em Advances in Data Analysis and Classification}, 5(4):281--300.
	
	\bibitem[Cheng et~al., 2016]{Che16}
	Cheng, W., Dryden, L.~L., and Huang, X.~Z. (2016).
	\newblock Bayesian registration of functions and curves.
	\newblock {\em Bayesian Analysis}, 11(2):447--475.
	
	\bibitem[Chiou and Li, 2007]{Chiou07}
	Chiou, J.~M. and Li, P.~L. (2007).
	\newblock Functional clustering and identifying substructures of longitudinal
	data.
	\newblock {\em J. R. Statist. Soc. B}, 69(4):679--699.
	
	\bibitem[Delaigle and Hall, 2010]{Del10}
	Delaigle, A. and Hall, P. (2010).
	\newblock Defining probability density for a distribution of random functions.
	\newblock {\em The Annals of Statistics}, 38:1171--1193.
	
	\bibitem[Ferraty and Vieu, 2006]{Fer06}
	Ferraty, F. and Vieu, P. (2006).
	\newblock {\em Nonparametric functional data analysis}.
	\newblock Springer Series in Statistics, Springer, New York.
	
	\bibitem[Gervini and Gasser, 2004]{Ger04}
	Gervini, D. and Gasser, T. (2004).
	\newblock Self-modelling warping functions.
	\newblock {\em J.R.Statist.Soc.B}, 66(4):959--971.
	
	\bibitem[Gower, 1975]{Gower:1975}
	Gower, J.~C. (1975).
	\newblock Generalized procrustes analysis.
	\newblock {\em Psychometrika}, 40(1):33--51.
	
	\bibitem[Green and Richardson, 2000]{Gre:00}
	Green, P.~J. and Richardson, S. (2000).
	\newblock Spatially corrrelated allocation models for count data.
	
	\bibitem[Hartigan and Wong, 1978]{Hart78}
	Hartigan, J.~A. and Wong, M.~A. (1978).
	\newblock Algorithm as 1326: A k-means clustering algorithm.
	\newblock {\em Applied Statistics}, 28:100--108.
	
	\bibitem[Hubert and Arabie, 1985]{Hub:85}
	Hubert, L. and Arabie, P. (1985).
	\newblock Comparing partitions.
	\newblock {\em J. Classif.}, 2:193--218.
	
	\bibitem[Ieva et~al., 2013]{Iev13}
	Ieva, F., Paganoni, A.~M., Pigoli, D., and Vitelli, V. (2013).
	\newblock Multivariate functional clustering for the analysis of ecg curves
	morphology.
	\newblock {\em Journal of the Royal Statistical Society. Series C},
	62(3):401--418.
	
	\bibitem[James and Sugar, 2003]{Jam03}
	James, G.~M. and Sugar, C.~A. (2003).
	\newblock Clustering for sparsely sampled functional data.
	\newblock {\em Journal of the American Statistical Association},
	98(462):397--408.
	
	\bibitem[Kenneth and David, 2004]{Ken04}
	Kenneth, P.~B. and David, R.~A. (2004).
	\newblock Understanding aic and bic in model selection.
	\newblock {\em Sociological Methods Research}, 33:261--304.
	
	\bibitem[Kim et~al., 2017]{Kim17}
	Kim, W.-S., Zeng, P., Shi, J.~Q., Lee, Y., and Paik, N.-J. (2017).
	\newblock Automatic tracking of hyoid bone motion from videofluoroscopic
	swallowing study with automatic smoothing and segmentation.
	\newblock {\em PLOS ONE}, page (to appear).
	
	\bibitem[Liu and Yang, 2009]{Liu:09}
	Liu, X. and Yang, M.~C.~K. (2009).
	\newblock Simultaneous curve registration and clustering for functional data.
	\newblock {\em Computational Statistics and Data Analysis}, 53:1361--1376.
	
	\bibitem[Lloyd, 1982]{Lloyd82}
	Lloyd, S.~P. (1982).
	\newblock Least squares quantization in pcm.
	\newblock {\em IEEE Transactions on Information Theory}, 28(2):129--137.
	
	\bibitem[MacQueen, 1967]{Mac67}
	MacQueen, J.~B. (1967).
	\newblock Some methods for classification and analysis of multivariate
	observations.
	\newblock {\em Proceedings of 5th Berkeley Symposium on Mathematical Statistics
		and Probability. University of California Press.}, pages 281--297.
	
	\bibitem[Marron et~al., 2015]{Mar15}
	Marron, J., Ramsay, J.~O., Sangalli, L.~M., and Srivastava, A. (2015).
	\newblock Functional data analysis of amplitude and phase variation.
	\newblock {\em Statistical Science}, 30(4):468--484.
	
	\bibitem[Raket, 2016]{Raket}
	Raket, L.~L. (2016).
	\newblock pavpop version 0.10.
	\newblock Available at \url{http://github.com/larslau/pavpop/}.
	
	\bibitem[Raket et~al., 2016]{Lars:2016}
	Raket, L.~L., Grimme, B., Schoner, G., Igel, C., and Markussen, B. (2016).
	\newblock Separating timing, movement conditions and individual differences in
	the analysis of human movement.
	\newblock {\em PLoS Comput Biol}, 12(9).
	
	\bibitem[Ramsay and Silverman, 2005]{Ram05}
	Ramsay, J.~O. and Silverman, B.~W. (2005).
	\newblock {\em Functional Data Analysis}.
	\newblock Springer-Verlag New York, USA.
	
	\bibitem[Rand, 1971]{Ran:71}
	Rand, W.~M. (1971).
	\newblock Objective criteria for the evaluation of clustering methods.
	\newblock {\em J. Am. Stat. Assoc.}, 66:846--850.
	
	\bibitem[Sam et~al., 2011]{Sam11}
	Sam, A., Chamroukhi, F., Govaert, G., and Aknin, P. (2011).
	\newblock Model-based clustering and segmentation of time series with changes
	in regime.
	\newblock {\em Advances in Data Analysis and Classification}, 5(4):301--322.
	
	\bibitem[Sangalli et~al., 2009]{San09}
	Sangalli, L.~M., Secchi, P., Vantini, S., and Veneziani, A. (2009).
	\newblock A case study in exploratory functional data analysis: geometrical
	features of the internal carotid artery.
	\newblock {\em J.Amer.Statist.Assoc.}, 104:37--48.
	
	\bibitem[Sangalli et~al., 2010]{San10}
	Sangalli, L.~M., Secchi, P., Vantini, S., and Vitelli, V. (2010).
	\newblock k-mean alignment for curve clustering.
	\newblock {\em Computational Statistics \& Data Analysis}, 54(5):1219--1233.
	
	\bibitem[Shi and Choi, 2011]{ShiB11}
	Shi, J.~Q. and Choi, T. (2011).
	\newblock {\em Gaussian Process Regression Analysis for Functional Data}.
	\newblock CRC Press Taylor \& Francis Group.
	
	\bibitem[Shi et~al., 2005]{Shi:05}
	Shi, J.~Q., Murray-Smith, R., and Titterington, D.~M. (2005).
	\newblock Hierarchical gaussian process mixtures for regression.
	\newblock {\em Statistics and Computing}, 15:31--41.
	
	\bibitem[Shi and Wang, 2008]{Shi:08}
	Shi, J.~Q. and Wang, B. (2008).
	\newblock Curve prediction and clustering with mixtures of gaussian process
	functional regression models.
	\newblock {\em Stat Comput}, 18(3):267--283.
	
	\bibitem[Shi et~al., 2012]{Shi:12}
	Shi, J.~Q., Wang, B., Will, E.~J., and West, R.~M. (2012).
	\newblock Mixed-effect gaussian process functional regression models with
	application to dose-response curve prediction.
	\newblock {\em Statist. Med.}, 31:3165--3177.
	
	\bibitem[Srivastava et~al., 2011a]{Sri11a}
	Srivastava, A., Klassen, E., Joshi, S.~H., and Jermyn, I.~H. (2011a).
	\newblock Shape analysis of elastic curves in euclidean spaces.
	\newblock {\em IEEE Trans. Pattern.Anal.Mach.Intell}, 33(7):1415--1428.
	
	\bibitem[Srivastava et~al., 2011b]{Sri11b}
	Srivastava, A., Wu, W., Kurtek, S., Klassen, E., and Marron, J.~S. (2011b).
	\newblock Registration of functional data using the fisher-rao metric.
	\newblock {\em Preprint. Available at arXiv:1103.3817v2}.
	
	\bibitem[Sugiura and Nariaki, 1978]{Sug78}
	Sugiura and Nariaki (1978).
	\newblock Further analysis of the data by akaike's information criterion and
	the finite corrections.
	\newblock {\em Communications in Statistics, Theory and Methods}, A7:13--26.
	
	\bibitem[Tang and M{\"u}ller, 1998]{Tan08}
	Tang, R. and M{\"u}ller, H.~G. (1998).
	\newblock Pairwise curve synchronization for functional data.
	\newblock {\em Biometricka}, 95(4):875--889.
	
	\bibitem[Tarpey and Kinateder, 2003]{Tar03}
	Tarpey, T. and Kinateder, K.~J. (2003).
	\newblock Clustering functional data.
	\newblock {\em Journal of Classification}, 20(1):93--114.
	
	\bibitem[Tokushige et~al., 2007]{Tok07}
	Tokushige, S., Yadohisa, H., and Inada, K. (2007).
	\newblock Crisp and fuzzy k-means clustering algorithms for multivariate
	functional data.
	\newblock {\em Computational Statistics}, 22:1--16.
	
\end{thebibliography}
	\bibliographystyle{apalike}
	\nocite{Cou90}

\setcounter{figure}{0}  
\setcounter{equation}{0}
\setcounter{section}{0}
\renewcommand{\theequation}{A.\arabic{equation}}
\renewcommand{\thefigure}{A.\arabic{figure}}
\renewcommand{\thesection}{Appendix \arabic{section}}  


\newpage

\section*{Appendices}  	

\section{Derivation of $M_{ki}$}
\label{ap:deri}
Using Bayes' theorem, the posterior distribution with respect to $\mf{z}$ has the form 
\[
p(\ve{z}|\ve{x}, \ve{\theta}, \ve{\beta}) \propto \prod_{i=1}^{N}\prod_{k=1}^{K}\pi_{ki}^{z_{ki}}p(\ve{x}_{i}|\ve{\theta}_{ki})^{z_{ki}}.
\]
By factorizing it over $i$, it is clear that the $\{\ve{z}_{i}, i = 1, \dots, N\}$ are independent under the posterior distribution. Hence,
\begin{equation}
\begin{split}
E(z_{ki}|\ve{x}, \ve{\theta}, \ve{\beta}) &= E(z_{ki}|\ve{x}_{i}, \ve{\theta}, \ve{\beta})\\
&= p(z_{ki}=1|\ve{x}_{i}, \ve{\theta}, \ve{\beta})\\
&= \frac{p(z_{ki}=1|\ve{\beta})p(\ve{x}_{i}|z_{ki} = 1, \ve{\theta})}{p(\ve{x}_{i}|\ve{\theta}, \ve{\beta})}\\
&= \frac{\pi_{ki}p(\ve{x}_{i}|\ve{\theta}_{ki})}{p(\ve{x}_{i}|\ve{\theta}, \ve{\beta})}.
\end{split}
\end{equation}
We know that
\begin{equation}
\begin{split}
p(\ve{x}_{i}|\ve{\theta}, \ve{\beta}) & = \sum_{\ve{z}_{i}}p(\ve{z}_{i}|\ve{\beta})p(\ve{x}_{i}|\ve{z}_{i}, \ve{\theta}, \ve{\beta})\\
& = \sum_{z_{ki}}\prod_{k=1}^{K}\big(p(z_{ki} = 1|\ve{\beta})p(\ve{x}_{i}|z_{ki}=1, \ve{\theta})\big)^{z_{ki}}\\
& = \sum_{j=1}^{K}\pi_{ji}p(\ve{x}_{i}|\ve{\theta}_{ji}).
\end{split}
\end{equation}
Thus,
\[
E(z_{ki}|\ve{x}, \ve{\theta}, \ve{\beta}) = \frac{\pi_{ki}p(\ve{x}_{i}|\ve{\theta}_{ki})}{\sum_{j=1}^{K}\pi_{ji}p(\ve{x}_{i}|\ve{\theta}_{ji})} .
\]

\section{Derivation of the linearized model}
\label{ap:linear}
At the linearized level, we do the first-order Taylor approximation of  model (\ref{eq:mixt3}) in Section \ref{subsec:esti} in the random warp $\ve{w}_{ki}$. We reconsider $g_{ki}(t)$ as a function with respect to $\ve{w}_{k}$ + $\ve{w}_{ki}$. Thus,  the linearization can be carried out around  the estimate of $\ve{w}_{k}$ plus $\ve{w}^{0}_{ki}$ obtained from the previous step. This results in a linear mixed-effects model as follows:
\[
\ve{x}_{ai}|_{z_{ki=1}} \approx \ve{x}_{ai}|_{z_{ki=1}, \ve{w}_{ki} = \ve{w}^{0}_{ki}} +  \nabla_{\ve{w}_{ki}}({\ve{x}_{ai}|_{z_{ki=1}}})|_{\ve{w}_{ki} = \ve{w}^{0}_{ki}}(\ve{w}_{ki} - \ve{w}^{0}_{ki})
\]
where
\begin{align*}
\ve{x}_{ai}|_{z_{ki=1}, \ve{w}_{ki} = \ve{w}^{0}_{ki}} &= \ve{\tau_{ak}}(g_{ki})|_{\ve{w}_{ki} = \ve{w}^{0}_{ki}} + \ve{r}_{aki} + \ve{\epsilon}\\
&=  \ve{\Psi}_{ki}|_{g_{ki} = g_{ki}^{0}}\ve{d}_{ak} + \ve{r}_{aki} + \ve{\epsilon}
\end{align*}
and according to the chain rule,
\begin{align*}
\nabla_{\ve{w}_{ki}}({\ve{x}_{ai}|_{z_{ki=1}}})|_{\ve{w}_{ki} = \ve{w}^{0}_{ki}}  & =  \Bigg\{\frac{\partial{\ve{x}_{ai}|_{z_{ki=1}, t = t_{j}}}}{\partial{g_{ki}}} \bigg(\nabla_{\ve{w}_{ki}}\big(g_{ki}(t_{j})\big)\bigg)^{\intercal}\Big|_{\ve{w}_{ki} = \ve{w}^{0}_{ki}}\Bigg\}_{j} \\                                                                                                                                   
& =    \bigg\{\frac{\partial \Big(\tau_{ak}\big(g_{ki}(t_{j})\big)\Big)}{\partial{g_{ki}}}\Big|_{g_{ki}= g_{ki}^{0}}\Big(\nabla_{\ve{w}_{ki}}\big(g_{ki}(t_{j})\big)\Big)^\intercal\Big|_{\ve{w}_{ki} = \ve{w}_{ki}^{0}}\bigg\}_{j} \in \ms{R}^{n_{i} \times n_{w}}   .                                                                              
\end{align*}
In practical, we use finite difference for calculating the derivative of $g_{ki}(t)$ with respect to $\ve{w}_{ki}$.

\section{Implementation of functional $k$-means}
\label{ap:imp}
We just use $d = ||\cdot||^{2}$ for the purpose of simplicity and convenience.
The iterative procedures  are as follows:
\begin{enumerate}
	\item Choose the initial values for the $z_{ki}$. We can use any clustering method for scalar variables $\{\ve{\beta}_{i}\}$ corresponding to the functional variables $\{\ve{x}_{i}\}$.
	\item Fix $z_{ki}$ and minimize $F$ with respect to the $\ve{\tau}_{ak}(g_{ki})$. In this phase, minimizing $F$ is equivalent to maximizing
	\[
	\sum_{k=1}^{K}\sum_{i=1}^{N}z_{ki}\bigg(\sum_{a=1}^{2}\big(\text{log}p(\ve{x}_{ai}|\ve{\theta}_{aki})\big)\bigg)
	\]
	with the assumption that the covariance matrix of $\ve{x}_{ai}$ are the same over all the subjects. The detailed estimation of $\{\ve{\theta}_{aki}\}$ have been mentioned before and the $\ve{\tau}_{ak}(g_{ki})$ can be obtained straightforwardly.
	\item Fix $\ve{\tau}_{ak}(g_{ki})$ and  minimize $F$ with respect to $z_{ki}$. Since the term $F$ in (\ref{eq:kmean}) involving different $i$ are independent, we can optimize $F$ for each $i$ separately by choosing $z_{ki}$ as follows
	\[
	z_{ki}  = \left\{
	\begin{array}{ll}
	1, \hspace{2mm}\text{if} \hspace{2mm}k = \argmin_{j}||\big(\ve{x}_{i} - \ve{\tau}_{j}(g_{ji})\big)||^{2};\\
	0, \hspace{2mm}\text{otherwise}.
	\end{array}
	\right.
	\]
	\item Repeat Step 2 and Step 3 until convergence.
\end{enumerate}

\section{Extra numerical results in simulation study}
\label{ap:sist}

\begin{figure}	
	\centering
	\begin{subfigure}[t]{\textwidth}
		\centering
		\includegraphics[width= 0.3\textwidth]{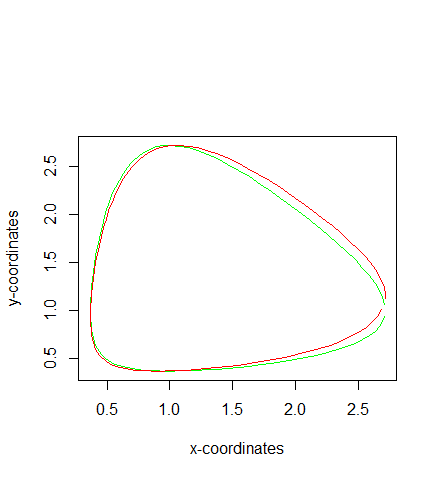}
		\includegraphics[width=0.3\textwidth]{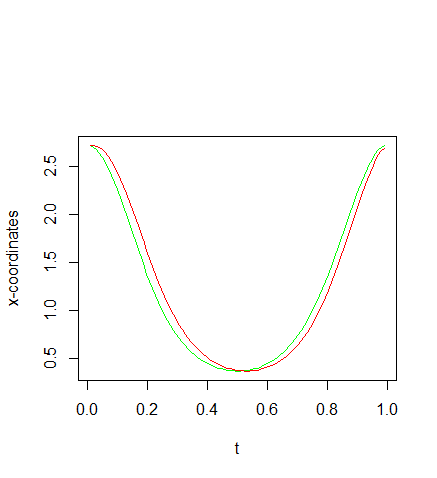}
		\includegraphics[width=0.3\textwidth]{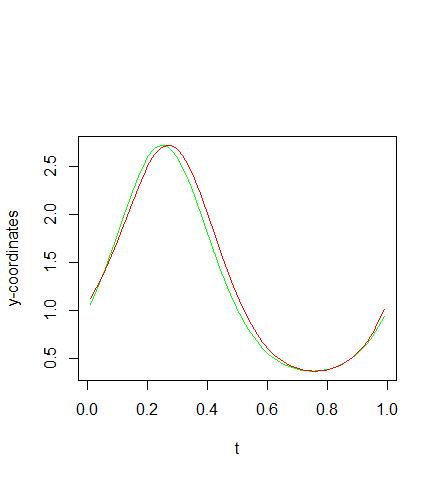}
		\caption{$b_{1} = 0.08$}		
	\end{subfigure}
	\quad
	\begin{subfigure}[t]{\textwidth}
		\centering
		\includegraphics[width= .3\textwidth]{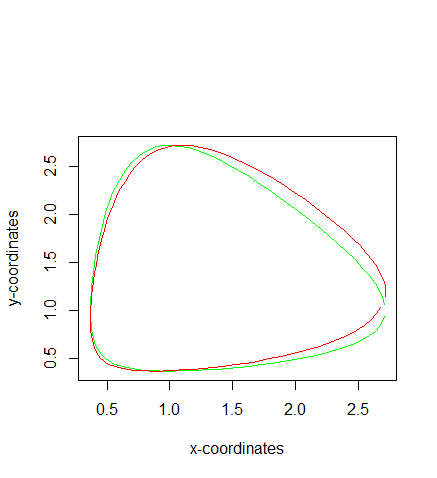}
		\includegraphics[width=.3\textwidth]{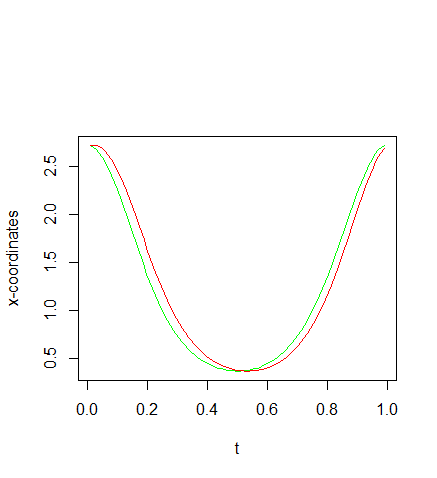}
		\includegraphics[width=.3\textwidth]{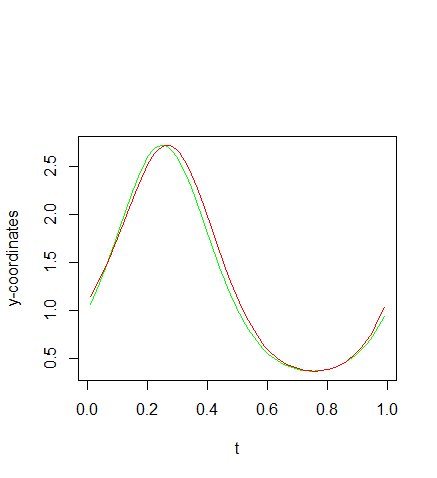}
		\caption{$b_{1} = 0.10$}		
	\end{subfigure}
	\quad
	\begin{subfigure}[t]{\textwidth}
		\centering
		\includegraphics[width= .3\textwidth]{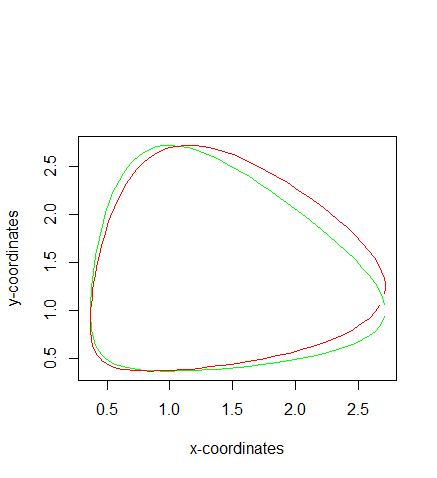}
		\includegraphics[width=.3\textwidth]{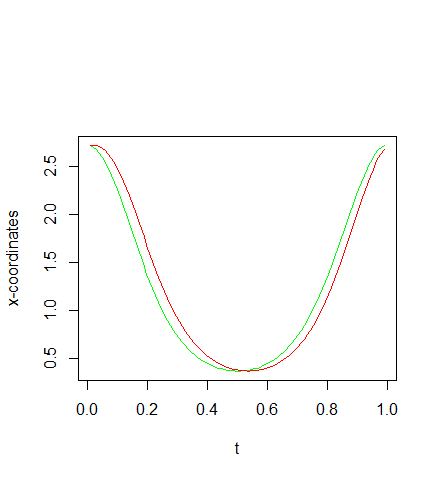}
		\includegraphics[width=.3\textwidth]{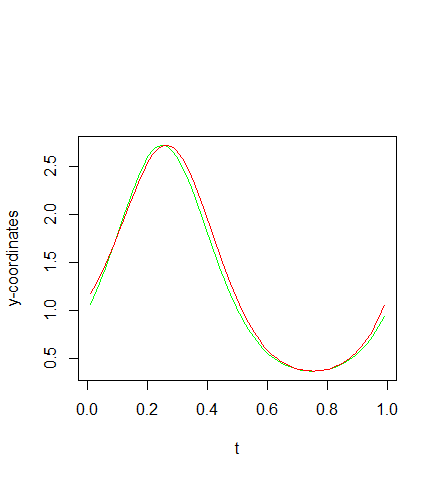}
		\caption{$b_{1} = 0.12$}		
	\end{subfigure}
	\caption{True mean curves $\bs{\mu}_{1}(t)$ (indicated by green lines) and $\bs{\mu}_{2}(t)$ (indicated by red lines) of group 1 and group 2 with $b_{1} = 0.08, 0.10, 0.12$.}\label{fig:cluster_true_mean_d010}
\end{figure}

\begin{figure}	
	\centering
	\begin{subfigure}[t]{0.45\textwidth}
		\centering
		\includegraphics[width= \textwidth]{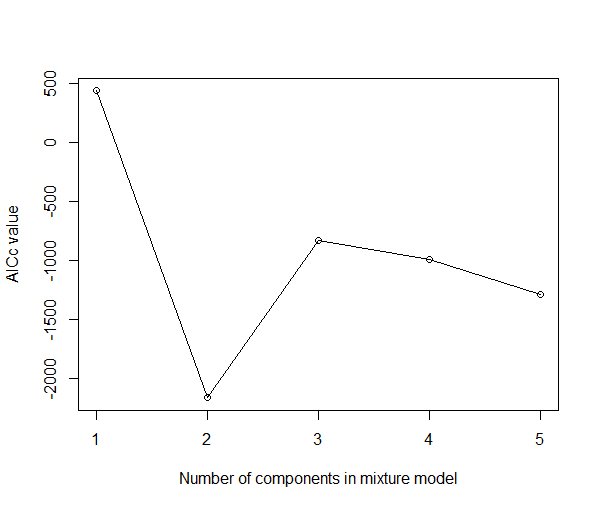}
		\caption{Scenario 1}		
	\end{subfigure}
	\quad
	\begin{subfigure}[t]{0.45\textwidth}
		\centering
		\includegraphics[width=\textwidth]{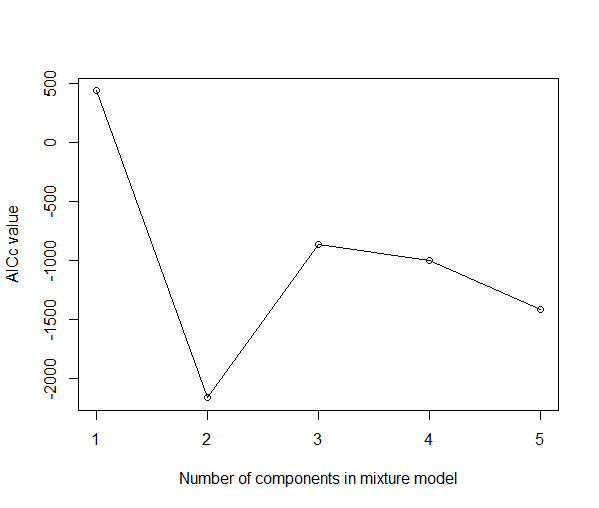}
		\caption{Scenario 2}
	\end{subfigure}
	\quad
	\begin{subfigure}[t]{0.45\textwidth}
		\centering
		\includegraphics[width=\textwidth]{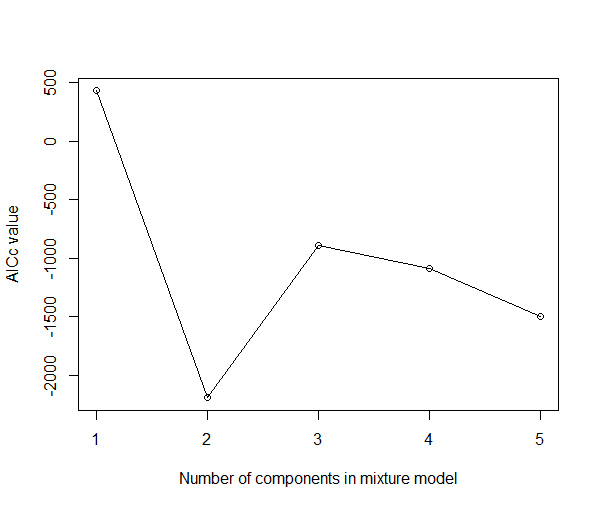}
		\caption{Scenario 3}
	\end{subfigure}
	\quad
	\begin{subfigure}[t]{0.45\textwidth}
		\centering
		\includegraphics[width=\textwidth]{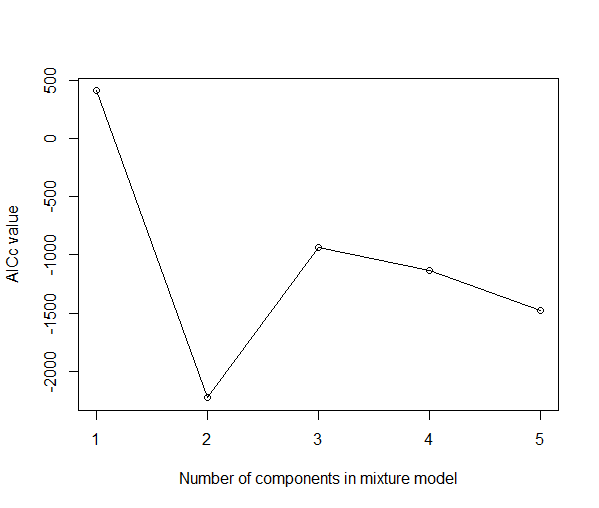}
		\caption{Scenario 4}
	\end{subfigure}
	\caption{The value of AICc calculated from one replication in each scenario for the method \textit{SRC}.}\label{fig:AICc}
\end{figure}

\begin{figure}	
	\centering
	\begin{subfigure}[t]{0.32\textwidth}
		\centering
		\includegraphics[width= 0.8\textwidth]{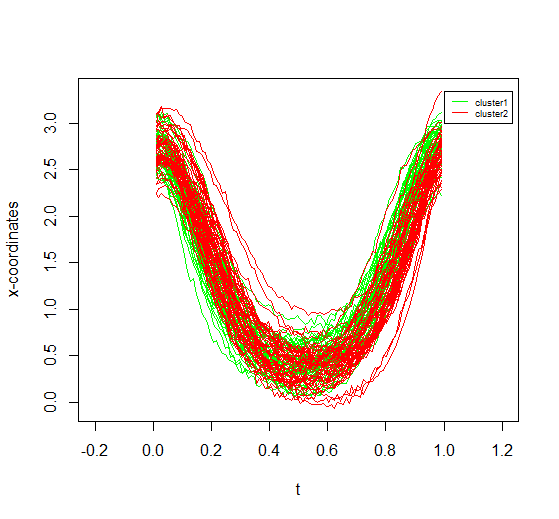}
		\caption{Original $\ve{x}_{1}(t)$}		
	\end{subfigure}
	\quad
	\begin{subfigure}[t]{0.32\textwidth}
		\centering
		\includegraphics[width= 0.8\textwidth]{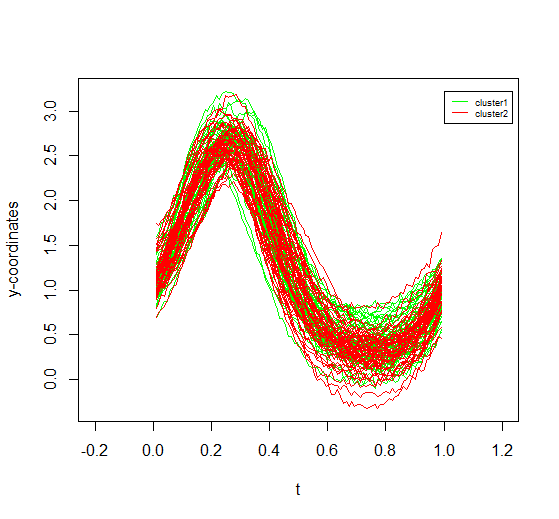}
		\caption{Original $\ve{x}_{2}(t)$}
	\end{subfigure}
	\quad
	\begin{subfigure}[t]{0.32\textwidth}
		\centering
		\includegraphics[width= 0.8\textwidth]{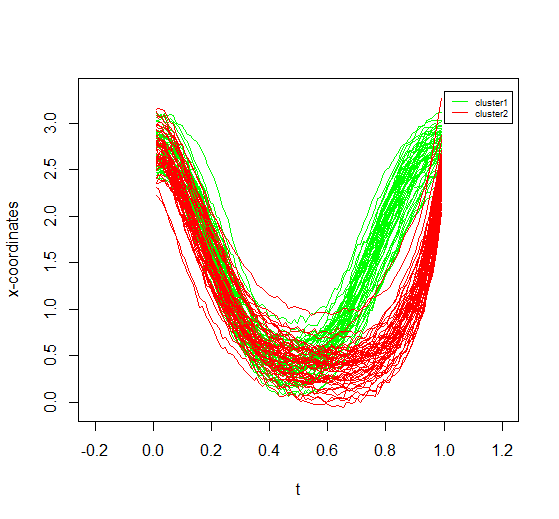}
		\caption{\textit{SRC}, aligned $\ve{x}_{1}(t)$}
	\end{subfigure}
	\quad
	\begin{subfigure}[t]{0.32\textwidth}
		\centering
		\includegraphics[width= 0.8\textwidth]{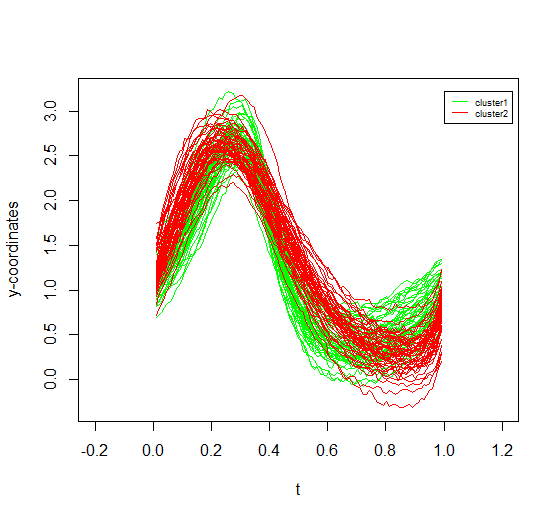}
		\caption{\textit{SRC}, aligned $\ve{x}_{2}(t)$}
	\end{subfigure}
	\quad
	\begin{subfigure}[t]{0.32\textwidth}
		\centering
		\includegraphics[width= 0.8\textwidth]{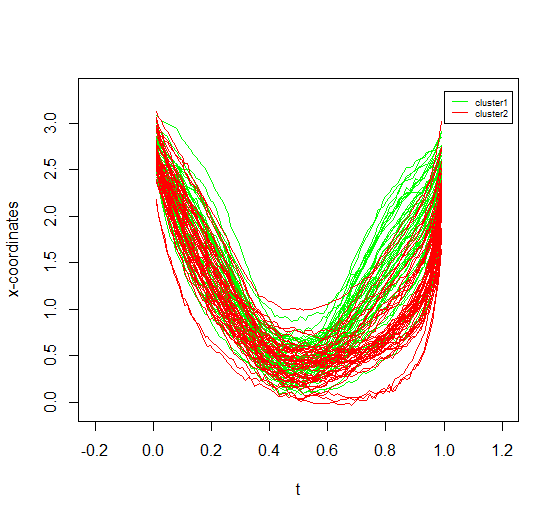}
		\caption{\textit{SRC-f}, aligned $\ve{x}_{1}(t)$}
	\end{subfigure}
	\quad
	\begin{subfigure}[t]{0.32\textwidth}
		\centering
		\includegraphics[width= 0.8\textwidth]{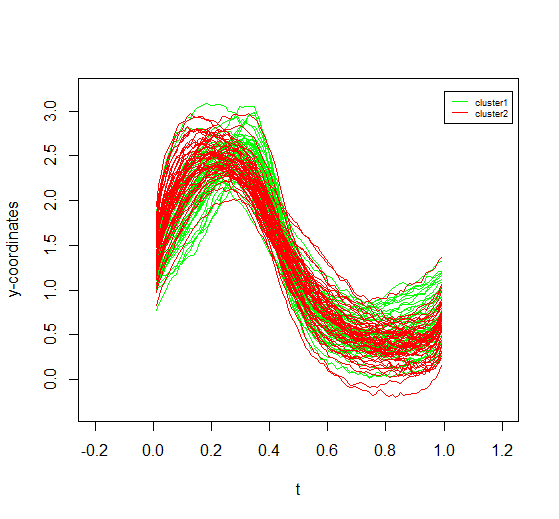}
		\caption{\textit{SRC-f}, aligned $\ve{x}_{2}(t)$}
	\end{subfigure}
	\quad
	\begin{subfigure}[t]{0.32\textwidth}
		\centering
		\includegraphics[width= 0.8\textwidth]{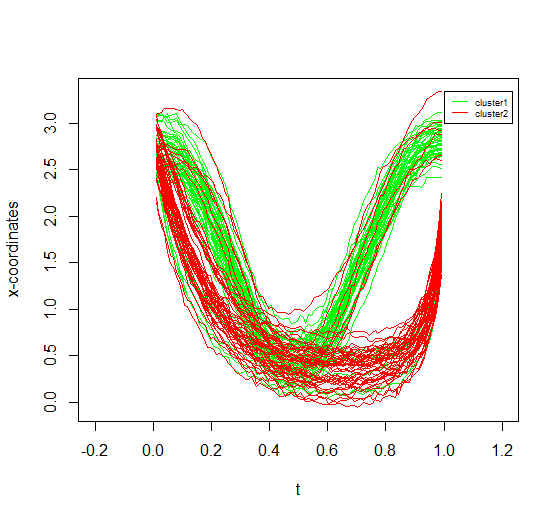}
		\caption{\textit{k-means-f}, aligned $\ve{x}_{1}(t)$}
	\end{subfigure}
	\quad
	\begin{subfigure}[t]{0.32\textwidth}
		\centering
		\includegraphics[width= 0.8\textwidth]{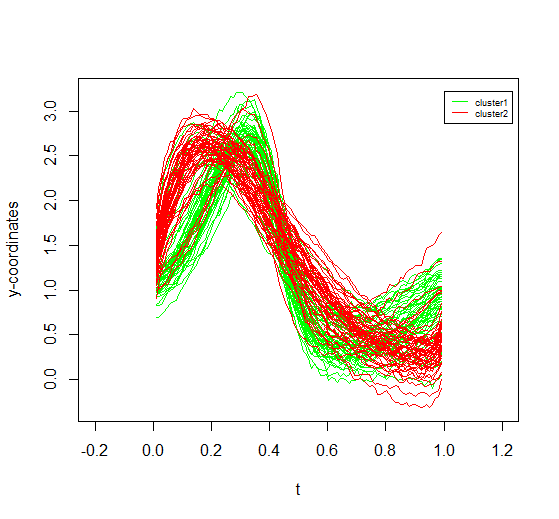}
		\caption{\textit{k-means-f}, aligned $\ve{x}_{2}(t)$}
	\end{subfigure}
	\caption{(a) and (b) are simulated 2D curves of two groups (green and red). (c)-(h) are aligned individual cruves by \textit{SRC}, \textit{SRC-f} and \textit{k-means-f}.}\label{fig:rec1}
\end{figure}

Figures \ref{fig:cluster_true_mean_d010} to \ref{fig:rec1} provide extra numerical results for the simulated examples discussed in Section \ref{sec:simu}. 

An extra example is given below.

\section{A simulation example in an extreme scenario}
\label{ap:exts}
\begin{figure}	
	\centering
	\begin{subfigure}[t]{0.32\textwidth}
		\centering
		\includegraphics[width= 0.8\textwidth]{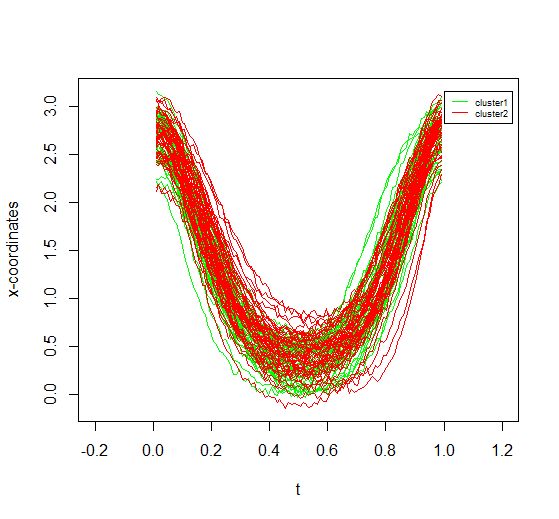}
		\caption{Original $\ve{x}_{1}(t)$}		
	\end{subfigure}
	\quad
	\begin{subfigure}[t]{0.32\textwidth}
		\centering
		\includegraphics[width= 0.8\textwidth]{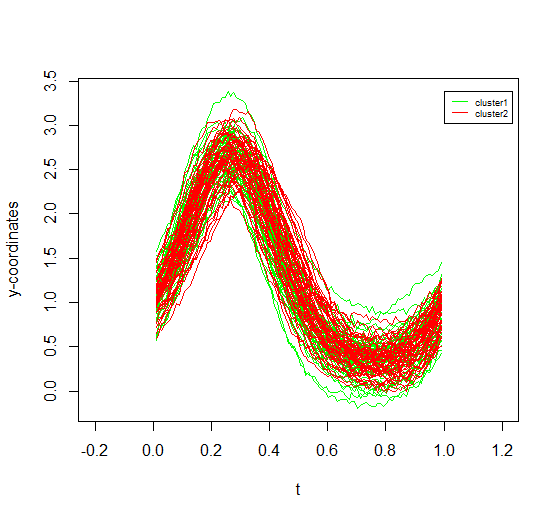}
		\caption{Original $\ve{x}_{2}(t)$}
	\end{subfigure}
	\quad
	\begin{subfigure}[t]{0.32\textwidth}
		\centering
		\includegraphics[width= 0.8\textwidth]{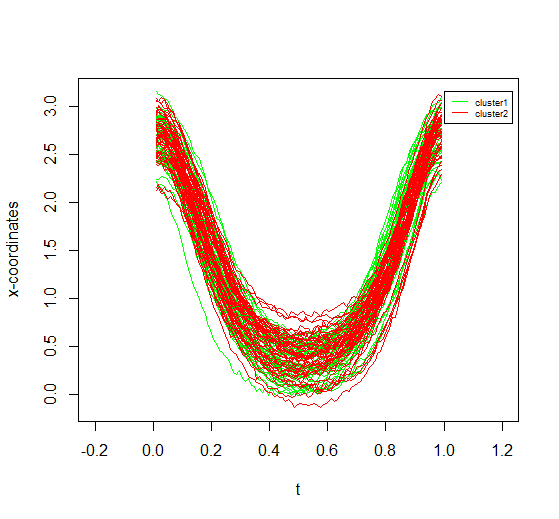}
		\caption{\textit{SRC}, aligned $\ve{x}_{1}(t)$}
	\end{subfigure}
	\quad
	\begin{subfigure}[t]{0.32\textwidth}
		\centering
		\includegraphics[width= 0.8\textwidth]{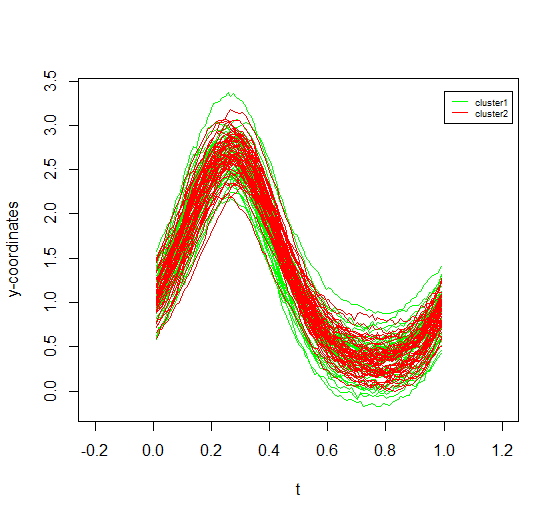}
		\caption{\textit{SRC}, aligned $\ve{x}_{2}(t)$}
	\end{subfigure}
	\quad
	\begin{subfigure}[t]{0.32\textwidth}
		\centering
		\includegraphics[width= 0.8\textwidth]{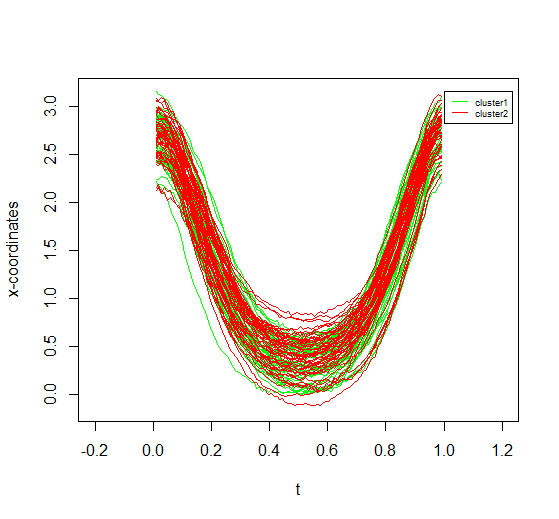}
		\caption{\textit{SRC-f}, aligned $\ve{x}_{1}(t)$}
	\end{subfigure}
	\quad
	\begin{subfigure}[t]{0.32\textwidth}
		\centering
		\includegraphics[width= 0.8\textwidth]{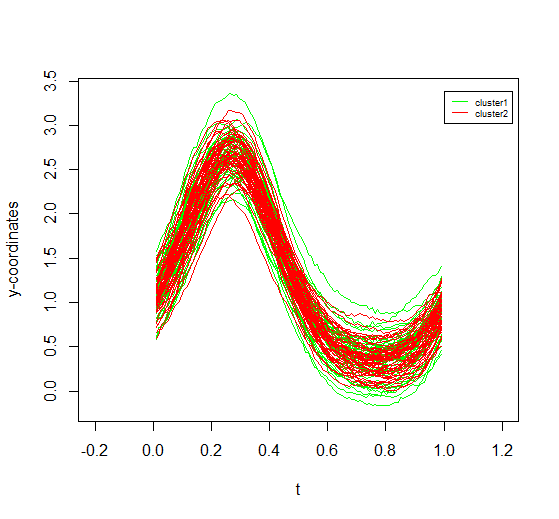}
		\caption{\textit{SRC-f}, aligned $\ve{x}_{2}(t)$}
	\end{subfigure}
	\quad
	\begin{subfigure}[t]{0.32\textwidth}
		\centering
		\includegraphics[width= 0.8\textwidth]{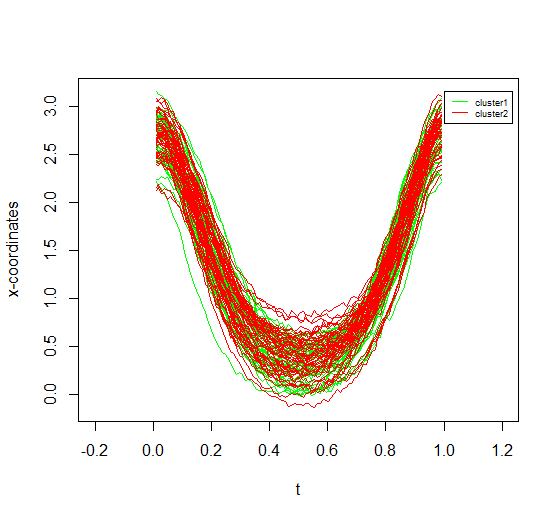}
		\caption{\textit{k-means-f}, aligned $\ve{x}_{1}(t)$}
	\end{subfigure}
	\quad
	\begin{subfigure}[t]{0.32\textwidth}
		\centering
		\includegraphics[width= 0.8\textwidth]{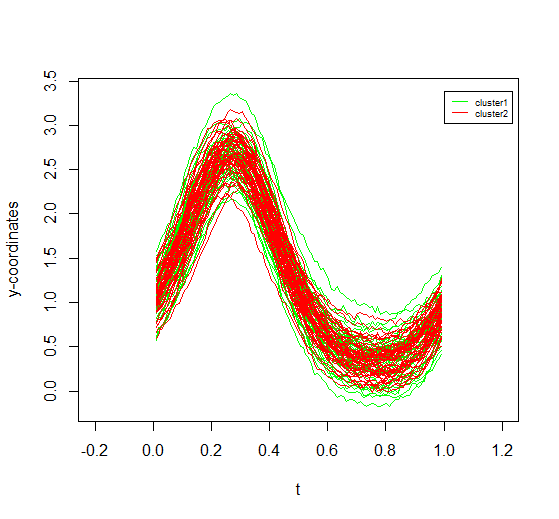}
		\caption{\textit{k-means-f}, aligned $\ve{x}_{2}(t)$}
	\end{subfigure}
	\caption{(a) and (b) are simulated 2D curves of two groups (green and red). (c)-(h) are aligned individual cruves by \textit{SRC}, \textit{SRC-f} and \textit{k-means-f}.}\label{fig:rec01}
\end{figure}

\begin{figure}	
	\centering
	\begin{subfigure}[t]{0.45\textwidth}
		\centering
		\includegraphics[width= \textwidth]{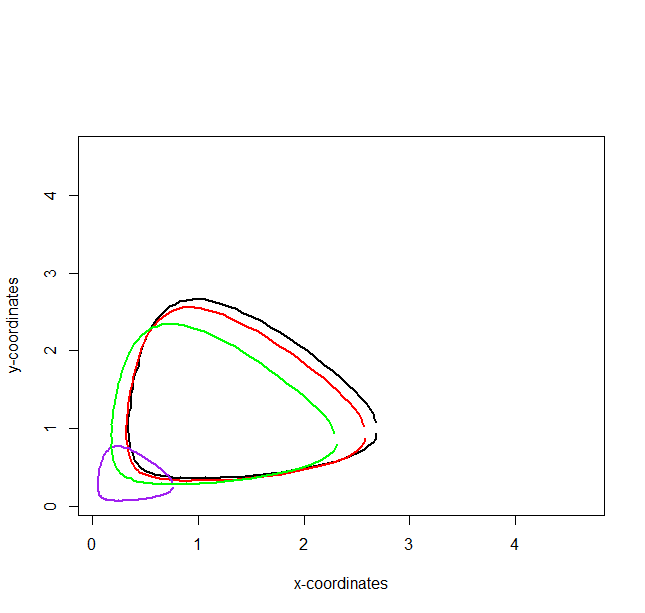}
		\caption{Cluster 1}		
	\end{subfigure}
	\quad
	\begin{subfigure}[t]{0.45\textwidth}
		\centering
		\includegraphics[width= \textwidth]{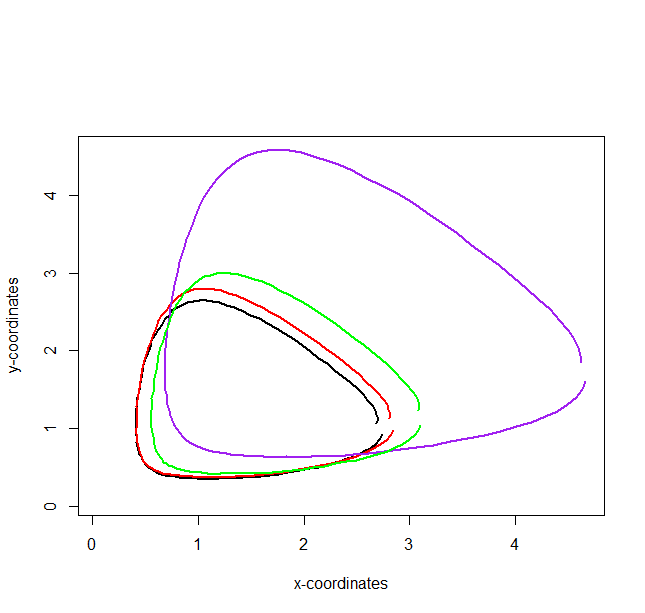}
		\caption{Cluster 2}		
	\end{subfigure}	
	\caption{Mean functions for 2D curves from two clusters. The \emph{\color{black}black lines} are true mean functions. The \emph{red lines}, \emph{purple lines} and \emph{green lines} are respectively corresponding to results obtained from the model \textit{SRC}, \textit{SRC-f} and \textit{k-means-f}.}\label{fig:rec02}
\end{figure}

It is not uncommon that sometimes the functional variables  provide  little information so that it fails to implement the clustering just based on those curves. However, the addition of scalar variables can make the clustering possible. We simulate a run of $N = 100$ (sample size), with $4\sigma_{w}^{2} = \sigma_{r}^{2} = \sigma^{2} = 0.02^2, b_{1} = 0.05, b_{2} = 0.8$, and $N_{1} = N_{2} = 50$ curves  in each group. Figure \ref{fig:rec01} displays the individual aligned curves resulting from three methods, from which no discernible clusters are visiable. The  RI and ARI for \textit{SRC}, \textit{SRC-f} and \textit{k-means-f} are, however, markedly different with the values of $(0.82, 0.50, 0.50)$ and $(0.64, 0, 0)$ respectively. Figure \ref{fig:rec02} summarizes the mean functions of two clusters by three methods. Their values of \textit{rase} are 1.1, 18.3 and 4.3  respectively. Those results show that the use of SRC leads to meaningful findings but the other twos are equivalent to random guess.

\section{Extra numerical results in real data analysis} \label{sec:appreal}

\begin{figure}[ht!]
	\centering
	\includegraphics[width = 0.55\textwidth]{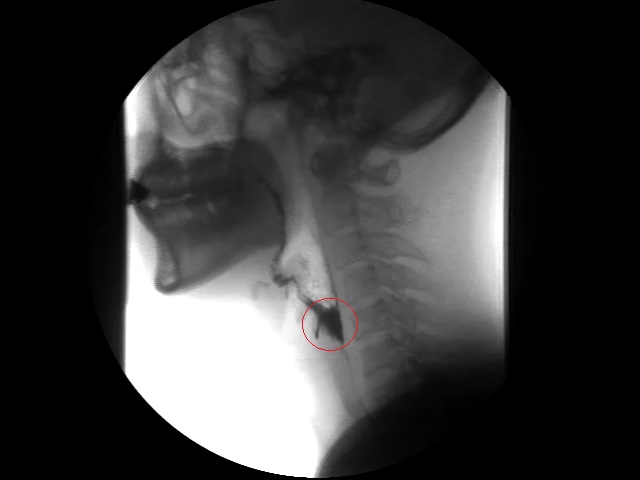}
	\caption{Highlight of Pyriform Sinus Residue, covered by the red circle}\label{fig:pyriform}
\end{figure}

\begin{figure}[ht!]
	\centering
	\includegraphics[width = 0.65\textwidth]{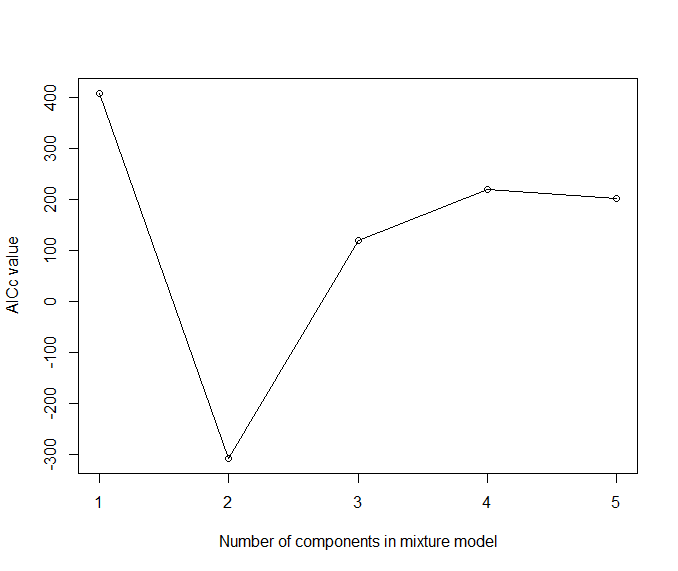}
	\caption{The values of AICc for $SRC$}\label{fig:AICc_real1}
\end{figure}

\begin{figure}	
	\centering
	\begin{subfigure}[t]{\textwidth}
		\centering
		\includegraphics[width= 0.3\textwidth]{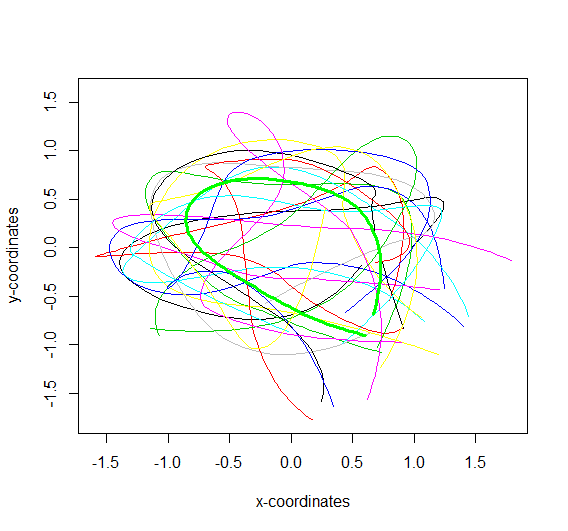}
		\includegraphics[width=0.3\textwidth]{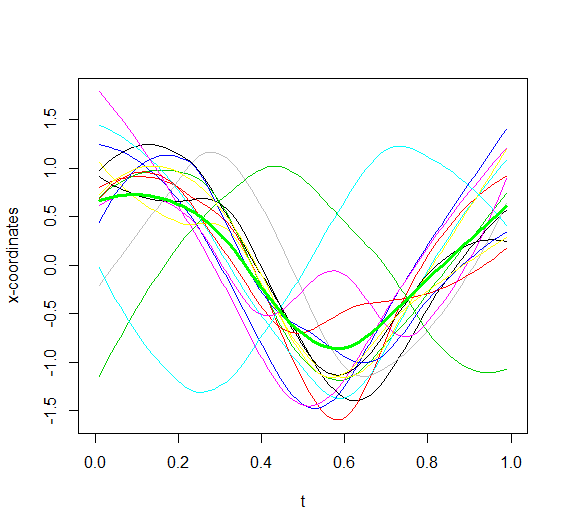}
		\includegraphics[width=0.3\textwidth]{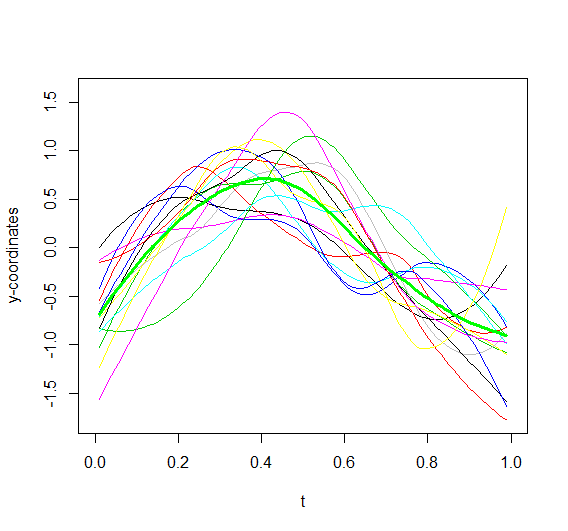}
		\caption{Curves from 15 normal people}		
	\end{subfigure}
	\quad
	\begin{subfigure}[t]{\textwidth}
		\centering
		\includegraphics[width= .3\textwidth]{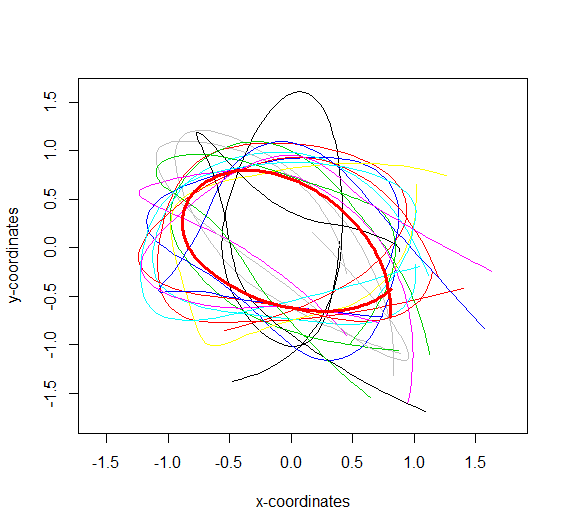}
		\includegraphics[width=.3\textwidth]{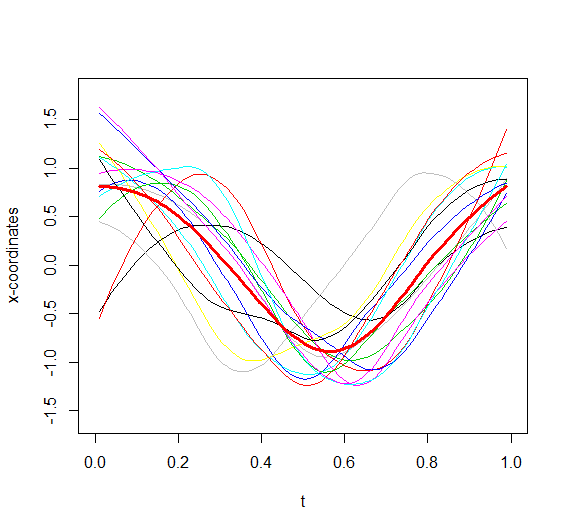}
		\includegraphics[width=.3\textwidth]{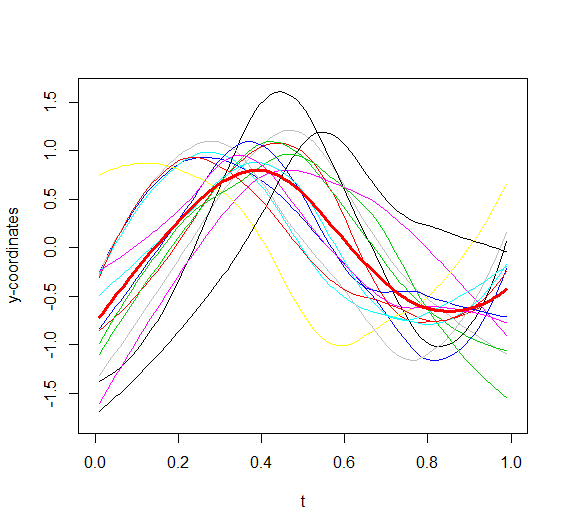}
		\caption{Curves from 15 abnormal people}		
	\end{subfigure}
	\caption{Curves of hyoid bone motion for two true groups, where  the bold curves in green (upper panel) and in red (lower panel) are the average mean curve for each group}\label{fig:real0}
\end{figure}

\begin{figure}	
	\centering
	\begin{subfigure}[t]{\textwidth}
		\centering
		\includegraphics[width= 0.3\textwidth]{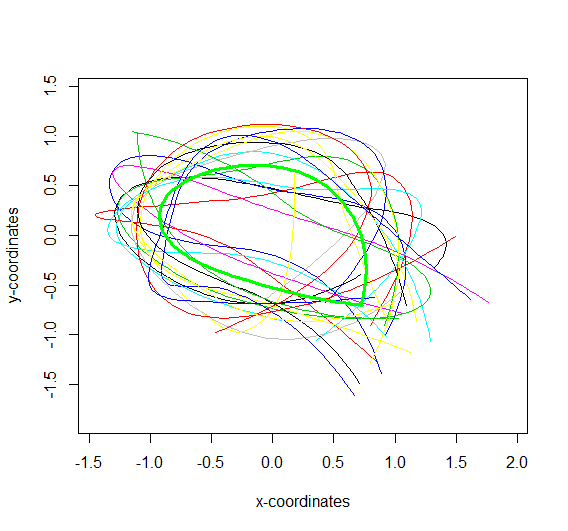}
		\includegraphics[width=0.3\textwidth]{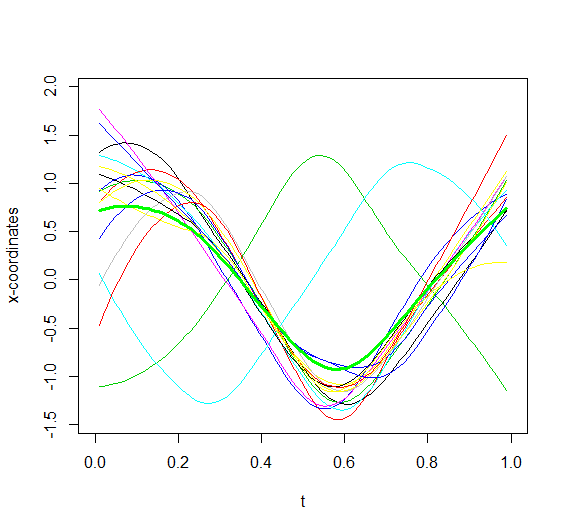}
		\includegraphics[width=0.3\textwidth]{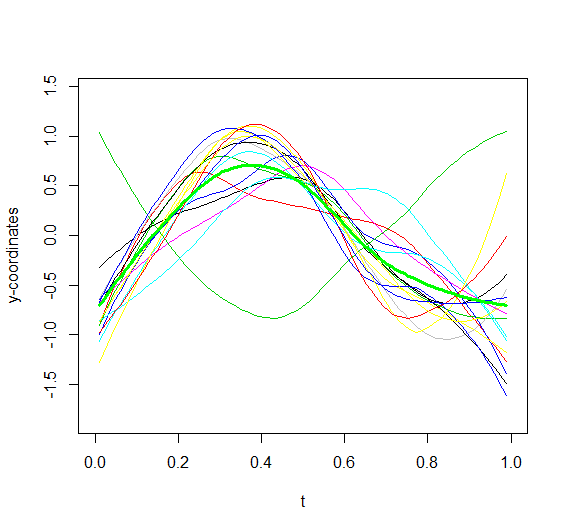}
		\caption{Cluster 1 with 16 people}		
	\end{subfigure}
	\quad
	\begin{subfigure}[t]{\textwidth}
		\centering
		\includegraphics[width= .3\textwidth]{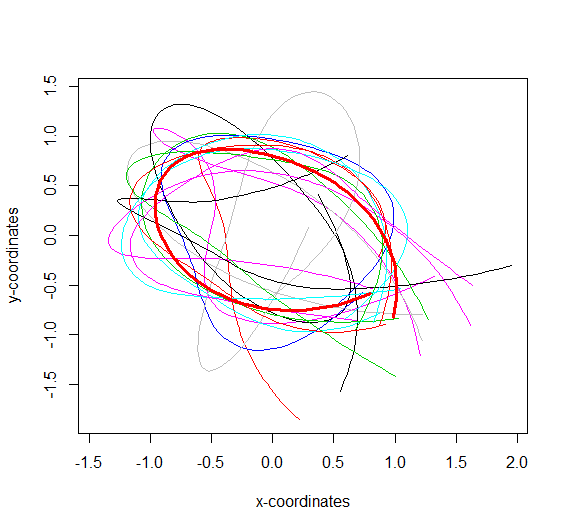}
		\includegraphics[width=.3\textwidth]{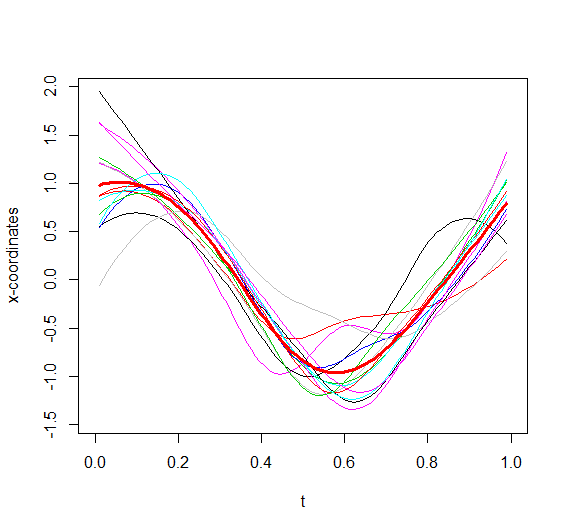}
		\includegraphics[width=.3\textwidth]{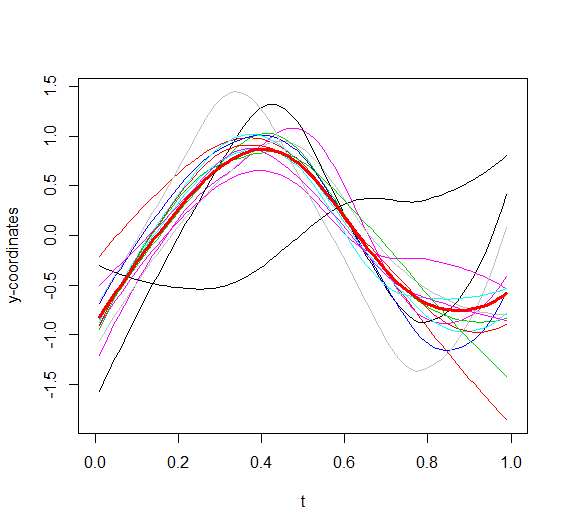}
		\caption{Cluster 2 with 14 people}		
	\end{subfigure}
	\caption{Curves of hyoid bone motion for two groups clustered by \textit{SRC}, where  the bold curves in green (upper panel) and in red (lower panel) are the average mean curve for each group}\label{fig:real1}
\end{figure}

Figures \ref{fig:pyriform} to \ref{fig:real1} provide extra numerical results for the real data example discussed in Section \ref{subsec:real}.

Figure \ref{fig:real1} shows the clustering results by SRC along with the registering effect, which are quite close to the true data shown in Figure \ref{fig:real0}.

\newpage

\label{ap:rean}

\end{document}